%% file: Arxiv_Final_Submission.tex
\documentclass[10pt,letterpaper]{article}
\usepackage[top=0.85in,left=1in,footskip=0.75in,marginparwidth=2in]{geometry}
\RequirePackage[OT1]{fontenc} 
\RequirePackage{graphicx, color,amsmath, amssymb, amsthm, amsfonts, subfig}
\RequirePackage{natbib}
\RequirePackage[colorlinks,citecolor=blue,urlcolor=blue]{hyperref}
\RequirePackage{mathpazo}
\RequirePackage{color,bm}
\RequirePackage{dsfont}
\RequirePackage{bigints}
\RequirePackage{enumerate}
\RequirePackage{bbm,lipsum}

\graphicspath{{./art/}}

\usepackage[linesnumbered,ruled,vlined]{algorithm2e}

\makeatletter
\renewcommand{\algocf@captiontext}[2]{#1\algocf@typo. \AlCapFnt{}#2} 
\def\@algocf@capt@plain{top}
\renewcommand{\algocf@makecaption}[2]{%
  \addtolength{\hsize}{\algomargin}%
  \sbox\@tempboxa{\algocf@captiontext{#1}{#2}}%
  \ifdim\wd\@tempboxa >\hsize
    \hskip .5\algomargin%
    \parbox[t]{\hsize}{\algocf@captiontext{#1}{#2}}
  \else%
    \global\@minipagefalse%
    \hbox to\hsize{\box\@tempboxa}
  \fi%
  \addtolength{\hsize}{-\algomargin}%
}
\makeatother

\newcommand\numberthis{\addtocounter{equation}{1}\tag{\theequation}}

\newcommand\eval[3]{{
  \left.\kern-\nulldelimiterspace 
  #1 
  \vphantom{\big|} 
  \right|_{#2}^{#3} 
  }}

\newcommand\restr[2]{{
  \left.\kern-\nulldelimiterspace 
  #1 
  \vphantom{\big|} 
  \right|_{#2} 
  }}

\theoremstyle{theorem}

\usepackage{hyperref} 

\newcommand{\R}{\mathds{R}}
\newcommand{\E}{\mathds{E}}
\newcommand{\N}{\mathds{N}}

\newcommand{\edr}{\mathrm{e}}
\newcommand{\ddr}{\mathrm{d}}
\newcommand{\indic}{\mathds{1}}

\def\simiid{\stackrel{\mbox{\scriptsize{iid}}}{\sim}}
\def\aceq{\stackrel{\mbox{\scriptsize{a.c.}}}{=}}
\def\eqd{\stackrel{\mbox{\scriptsize{d}}}{=}}

\newtheorem{definition}{Definition}
\newtheorem{theorem}{Theorem}
\newtheorem{proposition}{Proposition}
\newtheorem{corollary}{Corollary}
\newtheorem*{theorem*}{Theorem}
\newtheorem*{proposition*}{Proposition}

\input{definitions.tex}

\begin{document}

\vspace*{0.35in}

\begin{center}
{\Large
\begin{center}
\textbf{Restricted mean survival times for comparing grouped survival data: a Bayesian nonparametric approach.}
\end{center}
}
\bigskip
Alan Riva-Palacio$ ^{\,a}$, Antonio Lijoi$ ^{\,b}$ and 
Fabrizio Leisen$ ^{\,c}$
\\
\medskip
\begin{center}
\bf{ $ ^{a}$ IIMAS, Universidad Nacional Aut\'onoma de M\'exico, Mexico.}\\
\bf{ $ ^{b}$ Bocconi University, Italy.} \bf{ $ ^{c}$ King's College London, U.K.}
\end{center}
\medskip

\end{center}

\section*{Abstract}
Comparing survival experiences of different groups of data is an important issue in several applied problems. A typical example is where one wishes to investigate treatment effects. Here we propose a new Bayesian approach based on restricted mean survival times (RMST). A nonparametric prior is specified for the underlying survival functions: this extends the standard univariate neutral to the right processes to a multivariate setting and induces a prior for the RMST's. We rely on a representation as exponential functionals of compound subordinators to determine closed form expressions of prior and posterior mixed moments of RMST's. These results are used to approximate functionals of the posterior distribution of RMST's and are essential for  comparing time--to--event data arising from different samples.

\section{Introduction}

\noindent The analysis of time-to-event outcomes 
identifies a prominent area of statistical modeling and applications. 
Its relevance spans a wide range of disciplines such as, e.g.,  biology, medicine, epidemiology, engineering, economics, and sociology. Such a list is far from being exhaustive and 
keeps growing. Historically, survival analysis 
has spurred some of the most remarkable early developments of Bayesian nonparametric theory in the '70s. Seminal contributions in this direction are in \cite{doksum}, \cite{ferguson1974}, \cite{dykstralaud} and \cite{hjort}, just to mention a few. Most notably, such early Bayesian nonparametric literature heavily relied on the use of increasing L\'evy processes, namely processes with independent increments and almost surely increasing trajectories, to define priors in the infinite-dimensional space of survival or hazard functions. Targeting the survival or hazard functions is not surprising as they represent 
natural summaries of time-to-event data. The survival function contains all the information regarding the temporal profile of a study with an event time as endpoint. However, comparing populations with the survival function at a fixed time horizon, could make difficult the assessment of the survival experience. Specifically, when the survival curves across treatments cross each other at multiple points then it is not clear which treatment is to be preferred. As for the hazard ratio, which is widely used for comparing population subgroups, it may be misleading when the hazards 
cross each other in time, thus violating the standard proportionality assumption.

Given such limitations, here we consider an alternative approach, that compares different subgroups of time-to-event data in terms of the \textit{restricted mean survival time} (RMST). 
The RMST is the mean survival time of all subjects in the study population followed up to time horizon $t$, and it is simply the area under the survival curve up to $t$. See \cite{zhao2016} and  references therein. In the example of different treatment groups with crossing survival curves, the RMST gives a natural approach for population comparison in survival analysis where the highest RMST at a fixed time horizon $t$ will be preferred.

Additionally, when the proportional hazards assumption is violated, \cite{pepe} show that the RMST is more robust for comparing survival experiences of different population subgroups than other methods such as the logrank test. See also \cite{mantel} and \cite{peto}. Finally, unlike mean survival time, one can estimate the RMST even in the presence of censored observations. As a result, RMST has become a popular measure of treatment effect since it is intuitive and does not rely on strong distributional assumptions such as, e.g., proportionality of the hazards. The frequentist literature on the topic is quite rich: in addition to the previous papers one may refer to, e.g., \cite{roystonparmar}, \cite{zhao}, \cite{tian}, \cite{lee} and \cite{lawrence}. 

While the Bayesian approach is quite natural for addressing the problem of estimation of the RMST, up to our knowledge the literature is much more limited and we are aware of only two contributions that are somehow related to ours. One is in \cite{zhang}, which builds both on mixtures of the Dirichlet process as in \cite{antoniak} and on Dirichlet process mixtures as in \cite{lo} to define models that estimate the RMST. A similar approach is undertaken by \cite{Chen_et_al_BNP}, who use a mixture of covariate dependent stick--breaking processes to define a prior on the survival density: this induces a mixture prior also on the RMST. Though both papers have the merit of highlighting the potential benefits of Bayesian nonparametric (BNP) modeling for estimating the RMST, they are mostly focused on algorithmic and implementation aspects. Hence, a general theory on a wide class of BNP models that are suitably tailored to estimate RMST with data from multiple samples, and subject to a censoring mechanism, is still lacking. 

We fill this gap and develop a rigorous distribution theory for RMST, when the prior on the space of survival functions is the law of a vector of dependent neutral to the right processes. These are obtained through an exponential transformation of a vector $\bm{\xi}=\{\bm{\xi}(t)=(\xi_1(t),\ldots, \xi_d(t)):\: t\ge 0\}$ of dependent stochastic processes
\begin{equation}
	\label{eq:exp_funct}
\int_0^t\edr^{-\bm{\xi}(s)}\,\ddr s=\Big(\int_0^t \edr^{-\xi_1(s)}\,\ddr s,\ldots,\int_0^t \edr^{-\xi_d(s)}\,\ddr s\Big)
\end{equation}
where  $\xi_i(t)=\sum_{j\ge 1} W_{i,j} J_j\,\indic_{(0,t]}(Y_j)$, for each $i=1,\ldots,d$, and $\indic_A$ stands for the indicator function of set $A$. Here, $(W_{i,j})_{j\ge 1}$, for all $i=1,\ldots,d$, $(J_i)_{i\ge 1}$ and $(Y_i)_{i\ge 1}$ are independent sequences of independent positive random variables. The $i$--th component of the vector in \eqref{eq:exp_funct} is the RMST at $t$ for the $i$--th group of time--to--event outcomes. Our approach leads to novel results displaying multivariate mixed moments of the exponential functionals in \eqref{eq:exp_funct}, both \textit{a priori} and \textit{a posteriori} in closed form. As such, they can be evaluated either exactly or numerically. In particular, we use posterior mixed moments as a building block for estimating functionals of the posterior distribution of the RMST, as these are relevant for comparing survival experiences of different populations or samples. We do rely on a procedure that maximizes the Shannon entropy over the set of probability measures that are subject to $N$ moment constraints. Besides giving general results, we specialize them to the case of stratified compound random measure for which $W_{i,j}\in\{0,1\}$. This is an attractive specification if one is willing to single out shared and idiosyncratic features across different samples or subgroups of a population. 

The outline of the paper is as follows. In Section~2 we provide some background on compound random measures that are used to define the vector of exponential functionals in \eqref{eq:exp_funct}. In Section~3 we introduce multivariate neutral to the right priors and highlight their convenience for studying mean functionals. Results on RMST's are displayed in Section 4, while Section 5 provides detailed illustrations on simulated and real datasets. Finally, all proofs are deferred to the appendix together with further details on the numerical examples in Section~5.

\section{Compound random measures}\label{crm_sect}
The prior construction we consider heavily relies on the use of vectors of stochastic processes $\bm{\xi}=\{\bm{\xi}(t)=(\xi_1(t_1),\ldots,\xi_d(t)):\: t\ge 0\}$ such that
\begin{itemize}\addtolength{\itemsep}{0.4\baselineskip}
	\item[(i)]  for any $k\ge 2$ and  $0=t_0<t_1<\ldots<t_k$, the increments $\bm{\xi}(t_{1})-\bm{\xi}(t_0),\bm{\xi}(t_{2})-\bm{\xi}(t_1),\,\ldots,\, \bm{\xi}(t_{k})-\bm{\xi}(t_{k-1})$ are independent random vectors;
	\item[(ii)] $t\mapsto \xi_i(t)$ is almost surely increasing, for any $i=1,\ldots,d$. 
\end{itemize} 
Hence, each marginal process $\{\xi_i(t):\: t\ge 0\}$ has independent increments and is also known as an \textit{increasing additive process} according to the terminology in \cite{sato}. Under the additional assumptions of null deterministic drift and lack of fixed points of discontinuity, it is well--known that the Laplace transform of $\bm{\xi}$ can be represented as follows
\begin{equation}
	\label{eq:laplace_transform}
	\E\Big\{ \edr^{ -\bm{u}\cdot\bm{\xi}(t)} \Big\}=\edr^{-\psi_ t(\pmb{u})}=\edr^{-
	\int_{\re_+^d} ( 1-\edr^{-\bm{u}\cdot\bm{x}}) \,\nu_t(\d \pmb{x} )}
\end{equation}
for any $\bm{u}\in\R_+^d$, where $\bm{u}\,\cdot\,\bm{\xi}(t)=\sum_{i=1}^d u_i\xi_i(t)$ and $\bm{u}\cdot\bm{x}=\sum_{i=1}^d u_i x_i$ are the usual inner products. Moreover, $\{\nu_t:\: t\ge 0\}$ is a collection of measures on $\R_+^d$ such that $\nu_0\equiv 0$, for any $s<t$ one has $\nu_s(B)\le \nu_t(B)$, for any $B$, and  for any $t>0$
\[
\int_{\re_+^d}  \min \{ 1, \| \pmb{x} \| \} \nu_t ( \d \pmb{x} ) < \infty.
\] 
Henceforth, we refer to $\nu_t$ as the \textit{L\'evy measure} of $\bm{\xi}$. A well--known prior for the survival function in population $i$ is defined as the probability distribution of the random function
\begin{equation}
	\label{eq:ntr_marginal}
t\mapsto S_i(t)=\edr^{-\xi_i(t)}
\end{equation}
under the additional assumption that $\lim_{t\to+\infty}\xi_i(t)=+\infty$, almost surely, for every $i=1,\ldots,d$.  From \eqref{eq:laplace_transform} it is possible to deduce the Laplace transform of $\xi_i(t)$ by setting  $\bm{u}=(u_1,\ldots,u_d)$ with all $0$ entries but $u_i=u>0$, namely
\begin{equation}
	\label{eq:marginal_laplace}
	\E\Big\{\edr^{-u\xi_i(t)}\Big\}=\edr^{-\psi_{i,t}(u)}=\edr^{
	-\int_0^{+\infty}(1-\edr^{-ux})\,\nu_{i,t}(\ddr x)}.
\end{equation} 
The marginal L\'evy measure $\nu_{i,t}$ is obtained by marginalizing  $\nu_t$ with respect to all coordinates but the $i$--th one
\[
\nu_{i,t}(\ddr x_i)=\int_{\bm{x}_{-i}\in\R_+^{d-1}}\nu_t(\ddr \bm{x})
\]
with $\bm{x}_{-i}=(x_1,\ldots,x_{i-1},x_{i+1},\ldots,x_d)$. The definition in \eqref{eq:ntr_marginal} implies that the survival times in population $i$ have a \textit{neutral to the right} prior and we will use the notation $S_i\sim\mbox{NTR}(\xi_i)$.  It is further known that if $Y_1,\ldots,Y_n|S_i\simiid S_i$ and $S_i\sim\mbox{NTR}(\xi_i)$, then the posterior distribution of $S_i$, given $Y_1,\ldots,Y_n$, is still neutral to the right. Such a conjugacy property holds true even when given the data are right--censored. See \cite{doksum}. This is very useful as distributional results on functionals of $S_i$ \textit{a priori} extend also \textit{a posteriori}, with a suitable updating of $\xi_i$ and of the underlying L\'evy measure $\nu_t$. A notable example that will play a key role in the paper is the class of so--called compound random measures introduced in \cite{GL2017}.
\smallskip

\begin{definition}
A $d$--dimensional compound random measure on $\R_+$ is a vector of increasing additive processes $\bm{\xi}=\{\bm{\xi}(t):\: t\ge 0\}$ whose L\'evy measure equals
	\begin{equation*}
		\nu_t( \d \pmb{x} ) = \int z^{-d}h(x_1/z, \ldots , x_d/z) \nu_t^\star (\mathrm{d}z) \mathrm{d}\pmb{x} 
	\end{equation*}
	where $h$ is a multivariate probability density function that is referred to as score distribution, while $\nu^\star=\{\nu_t^\star:\: t\ge 0 \}$ is a collection of L\'evy measure on $\R_+^d$, also known as the directing L\'evy measures.
\end{definition}
\smallskip

Henceforth, we also use the notation $\bm{\xi}\sim\mbox{CoRM}_d(\nu^\star;h) $ to identify such a $d$--dimensional random measure. If $\nu_t^\star$ has Laplace exponent $\psi_t^\star$ then the conditional Laplace exponent of $\nu_t$, given $\pmb{W}\sim h$, is $\psi_t(\pmb{\lambda})= 
\psi_t^\star\left( \lambda_1 W_1 + \ldots + \lambda_d W_d  \right)  
$. This construction yields a straightforward and useful interpretation of $\bm{\xi}\sim\mbox{CoRM}_d(\nu^\star;h) $. Indeed, if $\{\xi^\star(t):\: t\ge 0\}$ is a an increasing additive process with L\'evy measures $\nu_t^\star$, as $t\ge 0$, it has the series representation 
\begin{equation}\label{eq:SerRep}
\xi^\star (t) = \sum_{i=1}^\infty J_i \indic_{(-\infty,t]) }(V_i),
\end{equation}
where $(J_i,V_i)$ are the points of a Poisson process with intensity $\nu^\star$, namely card$\{i\ge 1:\: (V_i,J_i)\in (0,t]\times A\}$ is Poisson distributed random variable with mean $\nu_t^\star(A)$, for any $A\subset \R_+\setminus [0,\epsilon)$ for $\epsilon>0$. If 
$\bm{W}_i=(W_{1,i},\ldots , W_{d,i})\simiid h$, then 
$\pmb{\xi}$ has series representation
$$
\pmb{\xi}(t) = 
\Big( \sum_{i=1}^\infty W_{1,i}J_i \mathds{1}_{ (-\infty,t]}(V_i), \ldots,  
\sum_{i=1}^\infty W_{d,i} J_i \mathds{1}_{ (-\infty,t]}(V_i) \Big).
$$
\noindent For further details on compound random measures see \cite{riva2021}. A specification that will be used throughout is where $\bm{\xi}$ is a stratified compound random measure. For the sake of simplicity, let $d=2$. If $\gamma>0$, $\alpha(\,\cdot\,)$ a probability density function on $\re_+$ and $s\mapsto \beta(s):=\int_s^\infty \alpha(u)\d u$, we specify $\xi^\star$ as a Log-Beta process $\xi^\star$, 
namely the directing L\'evy measure is
$$
\nu_t^\star( \d x) = \frac{\gamma}{1-\edr^{-x}}\:
\int_0^t \edr^{-\gamma \beta(s) x}\,\alpha(s )\:\d s\,\d x.
$$
On the other hand, we choose for the score distribution $\pmb{W}=(W_1,W_2)\sim h$ such that
\begin{align*}
	\pmb{W} = \begin{cases}
		(1,1)  & \text{ with probability } \pi_1
		\\
		(1,0)  & \text{ with probability } \pi_2
		\\
		(0,1)  & \text{ with probability } \pi_3
	\end{cases}
\end{align*}
for any $\pmb{\pi}=(\pi_1,\pi_2,\pi_3)\in\re_+^3$ such that $\pi_1+\pi_2+\pi_3=1$. The corresponding $2$--dimensional compound random measure is referred to as a $\pmb{\pi}$--compound Log-Beta process. 
This construction ensures a considerable degree of flexibility since one can achieve a wide range of dependence structures, ranging from degeneracy, namely $\xi_1\aceq\xi_2$, 
if $\pi_1=1$, to independence, $\xi_1\perp\xi_2$, if $\pi_1=0$ and $0<\pi_2<1$. In the former case, $\xi_i\aceq \xi^\star$, 
while in the latter the processes $\xi_1$ and $\xi_2$ almost surely have jumps at different times. 
An even more flexible construction is given by a stratification of the $\bm{\pi}$--compound Log-Beta process. Indeed, for $\bm{\pi}$--compound random measures weight dependence over time, see the series representation in \eqref{eq:SerRep}, can be too restrictive as it is assumed to be the same for all jump times. Accordingly, we let the score distribution $h_{t,\tau}$ of $\bm{W}=(W_1,W_2)$ depend on time and on $\tau$ as follows 
\begin{multline*}
	h_{t,\tau}=\Big\{\pi_{1,1}\indic_{\{t\le \tau\}}+\pi_{2,1}\indic_{\{t> \tau\}}\Big\}\,\delta_{(1,1)}+
	\Big\{\pi_{1,2}\indic_{\{t\le \tau\}}+\pi_{2,2}\indic_{\{t> \tau\}}\Big\}\,\delta_{(1,0)}\\[4pt]
	+\Big\{\pi_{1,3}\indic_{\{t\le \tau\}}+\pi_{2,3}\indic_{\{t> \tau\}}\Big\}\,\delta_{(0,1)}
\end{multline*}
where $\pi_{i,j}\ge 0$ and $|\bm{\pi}_i|:=\pi_{i,1}+\pi_{i,2}+\pi_{i,3}=1$ for each $i=1,2$. The corresponding vector of compound random measures is termed $(\pmb{\pi}_1,\pmb{\pi}_2)$--stratified-compound Log-Beta process. In the following section, such process will be used to define flexible priors on survival data.

\section{Multivariate neutral to the right priors}\label{sectNTRmultiv}

We will consider survival times associated with different samples or populations. This occurs when the data are recorded under different experimental conditions, such as survival times of patients who are diagnosed the same pathology and are treated in any of $d$ different medical centres or of patients who receive any of $d$ alternative treatments to be compared. In this case, we shall denote as $Y_{i,j}$ the $j$--th survival time in the $i$--th population. A natural dependence assumption for the array $\{(Y_{i,j})_{j\ge 1}:\: i=1,\ldots,d\}$ is partial exchangeability, namely
\begin{equation}
	\label{eq:part_exchange}
\{(Y_{i,j})_{j\ge 1}:\: i=1,\ldots,d\}\eqd \{(Y_{i,\pi_i(j)})_{j\ge 1}:\: i=1,\ldots,d\}
\end{equation}
for any choice of finite permutations $\pi_1,\ldots,\pi_d$ acting separately on the $d$ populations. This is equivalent to a condition of homogeneity within each population and heterogeneity across different populations, which is reasonable in the specific setting we are considering here. As a consequence of an obvious extension of de Finetti's representation theorem one has the following hierarchical structure
\begin{equation}
	\label{eq:partially_exchangeable_st}
\begin{split}
(Y_{i_1,j_1},\ldots,Y_{i_k,j_k})|(\tilde p_1,\ldots,\tilde p_d)\:
&\sim \tilde p_{i_1}\,\times\,\cdots\,\times\,\tilde p_{i_k}\\
(\tilde p_1,\ldots,\tilde p_d)&\sim Q_d
\end{split}
\end{equation}
for any choice of $k\ge 2$, $(i_1,\ldots,i_k)\in\{1,\ldots,d\}^k$ and $j_1,\ldots,j_k\ge 1$. The observations may also be right-censored, that is for censoring times $C_{i,j}$, which are assumed independent from the $Y_{i,j}$'s, the actual observations are $(T_{i,j},\Delta_{i,j})$, where $T_{i,j}=\min\{Y_{i,j},C_{i,j}\}:=Y_{i,j}\wedge C_{i,j}$ and $\Delta_{i,j}=1$ if $T_{i,j}=Y_{i,j}$ and $0$ otherwise. Henceforth, we assume that the $C_{i,j}$ are fixed. The distribution $Q_d$  of the vector $(\tilde p_1,\ldots,\tilde p_d)$ is the prior and in the paper will be specified as the probability distribution induced by $\bm{\xi}\sim\mbox{CoRM}(\nu^\star;h)$ through the following identity in distribution
\begin{equation}
	\label{eq:mult_ntr}
(\tilde p_1((t_1,+\infty)),\ldots,\tilde p_d((t_d,+\infty)))\eqd (\edr^{-\xi_1(t_1)},\ldots,\edr^{-\xi_d(t_d)}).
\end{equation}

\begin{definition}\label{def:multi_ntr}
A vector of random probability measures $(\tilde p_1,\ldots,\tilde p_d)$ on $\R_+$ defined as in \eqref{eq:mult_ntr} is termed $d$--dimensional neutral to the right prior governed by $\bm{\xi}$ and we use the notation $(\tilde p_1,\ldots,\tilde p_d)\sim\mbox{NTR}(\bm{\xi})$. If $\bm{\xi}$ is a $\bm{\pi}$(--stratified)--compound Log-Beta process, the associated NTR prior is referred to as $\bm{\pi}$(--stratified)--compound Beta-Stacy process.
\end{definition}
\smallskip

It follows that the marginal distributions of each $\tilde p_i$ in the above definition are neutral to the right, according to the definition in \cite{doksum}, and we use the notation $\tilde p_i\sim\mbox{NTR}(\xi_i)$. In the $\bm{\pi}$--compound Beta-Stacy case, the distribution of $\bm{W}_i=(W_{1,i},W_{2,i})$ tunes the so--called "\textit{borrowing of information}" across groups for such NTR prior. Indeed, when $\xi_1 \stackrel{\text{a.s.}}{=} \xi_2$ 
there is complete borrowing among groups since the distribution of the survival times is the same across groups. On the other hand if $\xi_1$ and $\xi_2$ are independent, then the survival times are unconditionally independent and there is no borrowing of information: observations from one group have no impact on inferences for the other group. These two extremes correspond to triplets $\bm{\pi}=(1,0,0)$ and $\bm{\pi}=(0,\pi_2,\pi_3)$, with $\pi_2,\pi_3>0$. Intermediate situations may be achieved by specifying suitable vectors $\bm{\pi}$: in particular, larger and smaller values of $\pi_1$ lead to situations that are closer to exchangeability and independence, respectively.

\begin{center}
\begin{figure}[h!]
\begin{center}
\begin{tabular}{ccc}
\includegraphics[width=0.3\linewidth,height=0.25\linewidth]{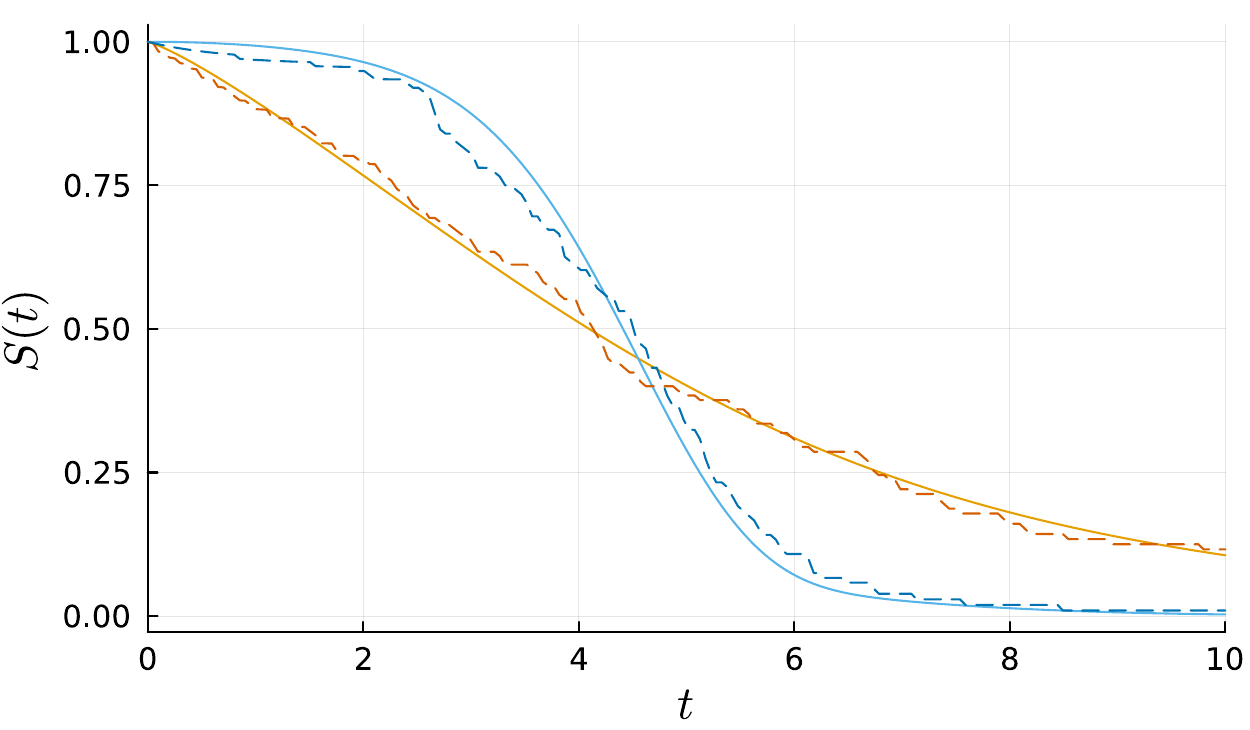} & \includegraphics[width=0.3\linewidth,height=0.25\linewidth]{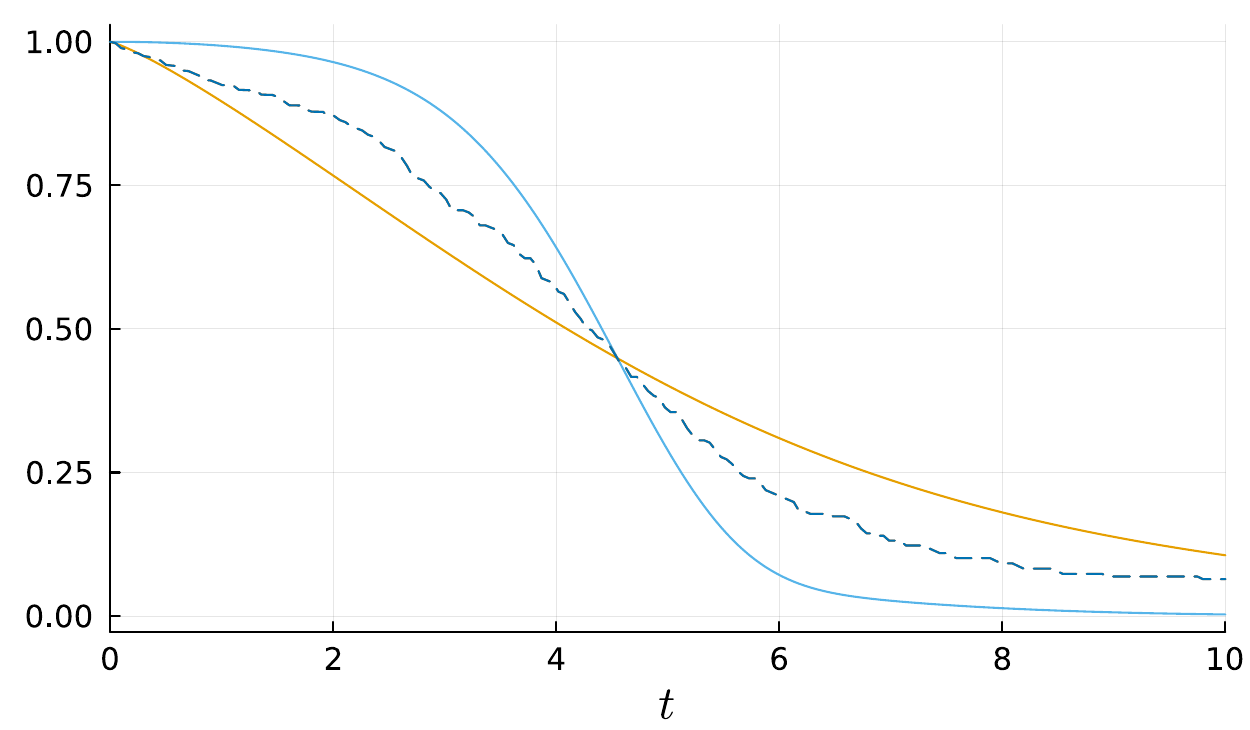} & \includegraphics[width=0.3\linewidth,height=0.25\linewidth]{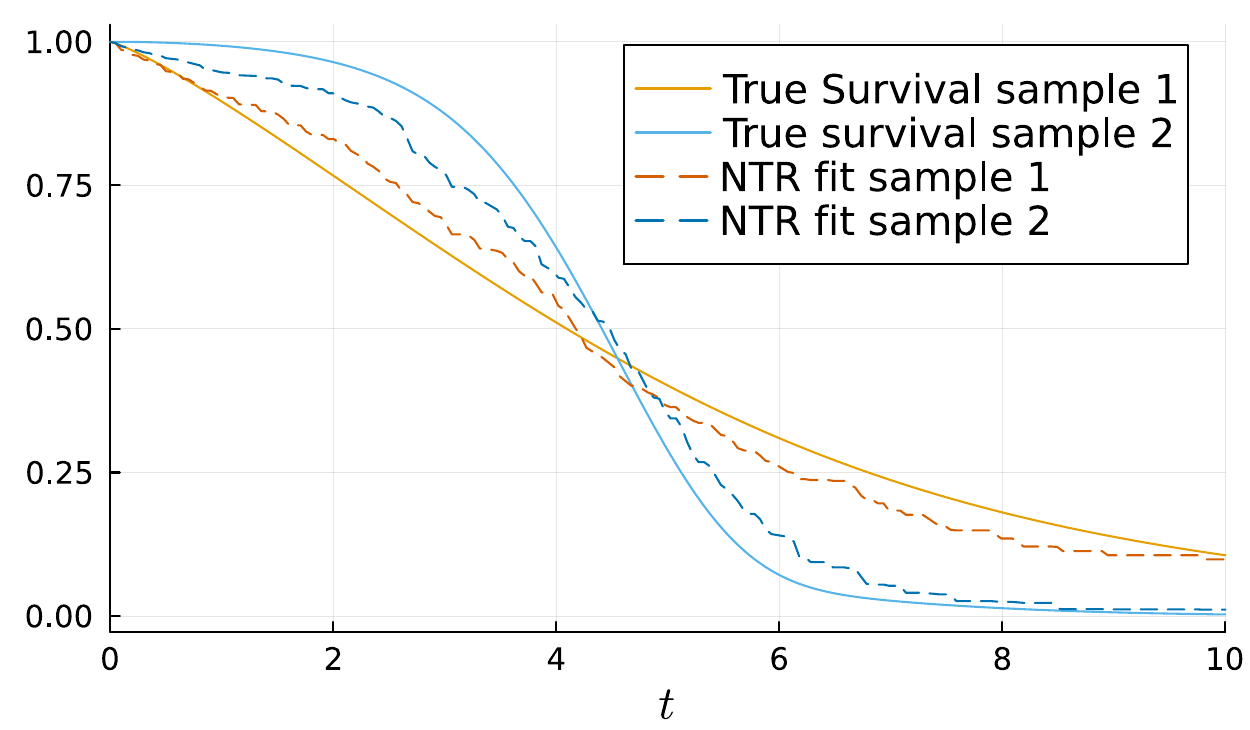}
 \\
(i) & (ii) & (iii)
\end{tabular}
\caption{Estimates of two survival functions with $\bm{\pi}$--compound Beta-Stacy priors: (i) $\pmb{\pi}=(0,0.5,0.5)$, (ii) $\pmb{\pi}=(1,0,0)$, (iii) $\pmb{\pi}=(0.5,0.25,0.25)$. The estimates are based on two independently generated datasets of size 150 each.  
			}\label{borrowfig}
\end{center}
\end{figure}
\end{center}

With the aim of visualizing the effects of different specifications of $\bm{\pi}$ on posterior inferences of such a prior specification, we consider a simple simulation scenario. The displayed posterior inferences have been evaluated by resorting to the theoretical background described in Section~\ref{AppB} in the Appendix, we highlight in particular that given possibly censored to the right data $\pmb{\mathcal{D}}$ the distribution of $\pmb{\xi}|\pmb{\mathcal{D}}$ is characterized in Theorem \ref{posteo} which enables posterior inference.
Time--to--event data were simulated independently both within and between the two samples, from two different distributions. In Figure~\ref{borrowfig} we reproduce plots of true data generating survival functions 
and their respective estimates for different choices of $\bm{\pi}$. In this case, it is not surprising that the best estimates are obtained under a model with $p_1=0$, while for the other cases with $p_1>0$ it is apparent that a shrinking of the two estimates occurs due to the effect of borrowing of information across the two samples.



\begin{center}
\begin{figure}[h]
\begin{center}
\begin{tabular}{cc}
\includegraphics[width=1.0\linewidth,height=0.25\linewidth]{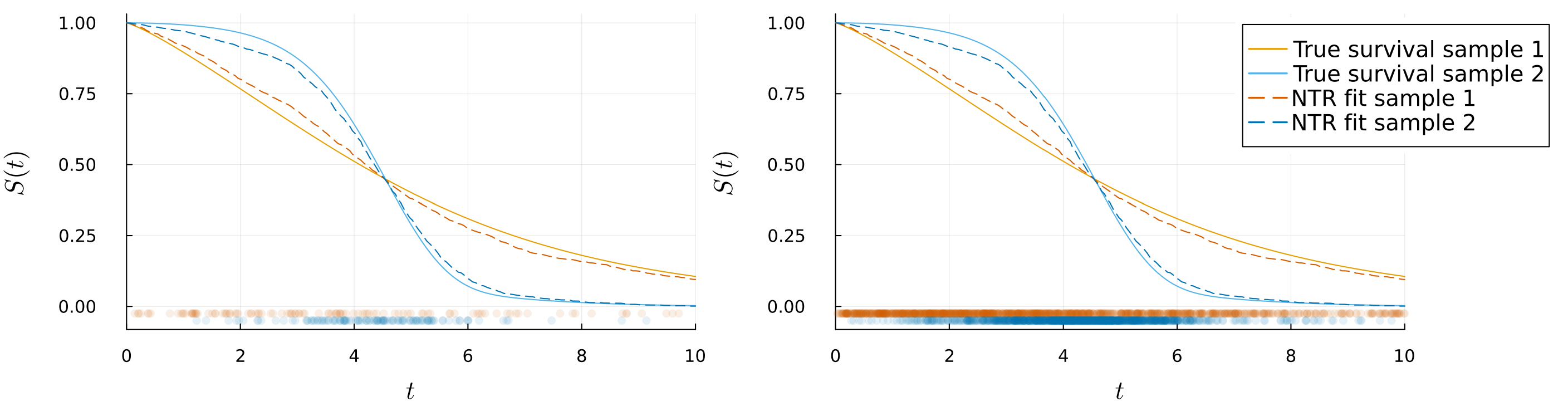} \\
(i) \quad\quad\quad\quad\quad\quad\quad\quad\quad\quad
\quad\quad\quad\quad\quad\quad\quad\quad\quad\quad(ii)\quad\quad \quad
\end{tabular}
\caption{Illustration of poor inference with un--stratified borrowing of information for NTR fits with $\bm{\pi}$--compound Beta-Stacy prior and increasing number of observations (i) $150$ and (ii) $1500$ in each population.}\label{fig:consistnostrat}
\end{center}
\end{figure}
\end{center}

\begin{center}
\begin{figure}[h!]
\begin{center}
\begin{tabular}{cc}
\includegraphics[width=1.0\linewidth,height=0.25\linewidth]{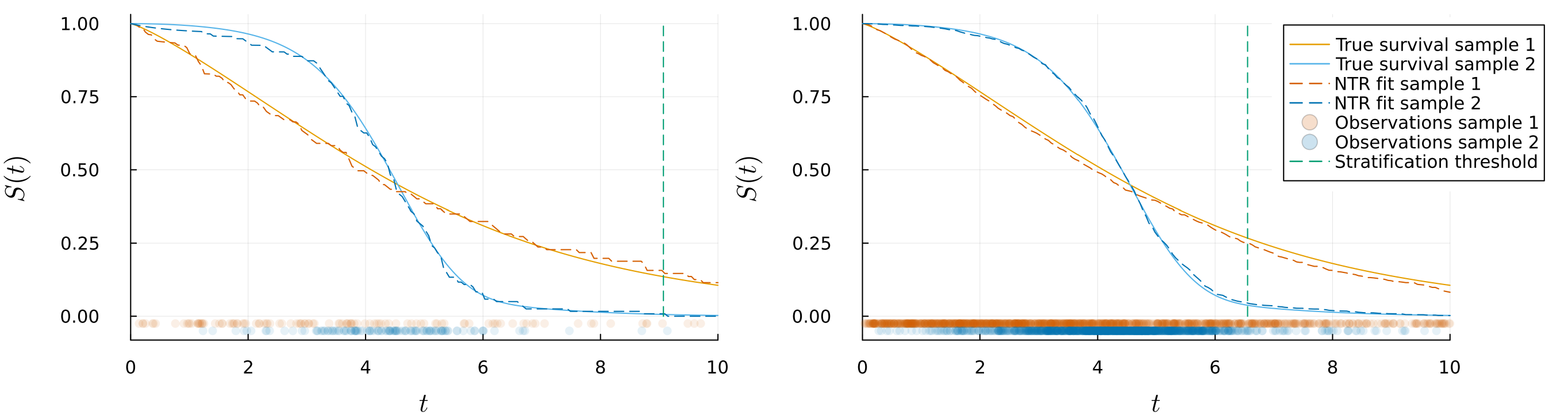} \\
(i) \quad\quad\quad\quad\quad\quad\quad\quad\quad\quad
\quad\quad\quad\quad\quad\quad\quad\quad\quad\quad(ii)\quad 
\end{tabular}
\caption{Illustration of flexible borrowing of information for NTR fits with $\bm{\pi}$--stratified-compound Beta-Stacy prior and increasing number of observations (i) $150$ and (ii) $1500$ in each population.}\label{fig:consiststrat}
	\end{center}
\end{figure}
\end{center}

Since $\bm{\pi}$--compound Beta-Stacy model may induce a strong shrinkage between estimates related to different samples, as in Figure~\ref{borrowfig} above, one may think of a different prior specification that stratifies data. Indeed, when the bulk of observations in one sample is below some threshold $\tau>0$ while a considerable amount of observations in the other sample is above $\tau$ borrowing of information is desirable only within the two clusters of observations: those above and those below the threshold $\tau$. This can be achieved through a $(\pmb{\pi}_1,\pmb{\pi}_2)$--stratified-compound Beta-Stacy prior. In Figures \ref{fig:consistnostrat} and \ref{fig:consiststrat} we illustrate the improvement of using the stratified construction when fitting survival data as detailed in Section~\ref{AppB} of the Appendix. 


\section{Restricted mean survival times}\label{sec:RMST}

In view of the prior specification in \eqref{eq:mult_ntr}, the mean survival time in population $i$ is $\mu_{i,\infty}=\int_0^{+\infty}\,\edr^{-\xi_i(s)}\,\ddr s$. These are also known as exponential functionals of an increasing additive process and have been extensively studied in the literature, for the case $d=1$, due to their relevance for financial applications. See, e.g., \cite{bertoinyor} and \cite{hirschyor}. In a Bayesian framework, and still with $d=1$, the investigation of distributional properties of $\mu_{i,\infty}$ has been thoroughly developed in \cite{epilijoiprunst}: here one can find conditions for the existence of $\mu_{i,\infty}$ and closed form expressions of the corresponding moments, of any order. Such moments are used to estimate the density of $\mu_{i,\infty}$ by resorting either ta a maximum-entropy (max-ent) algorithm or to a moment matching Ferguson and Klass algorithm. The setting in this paper differs from previous contributions in that we consider simultaneously $d\ge 2$ moments associated with different samples and populations. In view of the partial exchangeability assumption in \eqref{eq:part_exchange}, for any $i\ne j$ the survival times recorded in population $i$ impact also inferences on the mean of population $j$ and this induces a borrowing of information across different groups of data that appears reasonable in this framework.  Additionally, instead of considering $\mu_{i,\infty}$, we do consider restricted mean survival times and the $k$--th moment restricted to some interval $(0,t)$, namely 
\begin{equation}
	\label{eq:moments_ntr}
	\mu_{i,t}=\int_0^t \edr^{-\xi_i(s)}\,\ddr s,\qquad\mbox{ and } \qquad \mu_{i,t}^{(k)}=k\,\int_0^t s^{k-1}\edr^{-\xi_i(s)}\,\ddr s
\end{equation}
for each $i=1,2$. We will be further interested in evaluating mixed moments of the exponential integrals associated to $\xi_1$ and $\xi_2$ in \eqref{eq:moments_ntr} as they can be used for performing density estimation. Estimates of the densities will then be used for drawing inferences on the dissimilarity of survival experiences across groups. Important statistical quantities such as the variance, second central moment, skewness and kurtosis, which can be defined in terms of higher order standardize moments, can be considered to assess differences across groups. 

For survival analysis in particular the mean and variance of the survival time for different treatment groups in clinical trial studies can be of key importance for decision making. As an example we can consider a laboratory that will pursue drug development if for $v_1,v_2>0$ and $p_1,p_2\in[0,1]$
$$
\prob{ \mu_{\text{treatment}} - \mu_{\text{control}} > v_1 }\geq p_1,
\quad \prob{ \sigma^2_{\text{treatment}}  < v_2 }\geq p_2,
$$
by which not only the mean survival of the treatment is higher than the one for the control but also can be a set number of standard deviations away from it with high probability.
Let $\mathcal{D}$ denote possibly censored to the right observations in both groups and $c>0$.

One can estimate the posterior distribution of $(\mu_{1,t} - \mu_{2,t})^2$, conditionally on $\mathcal{D}$, using mixed moments of RMST's $(\mu_{1,t}^{(k)},\mu_{2,t}^{(k)})$ for $k=1,\ldots,K$. Such an estimate can be used to obtain an approximate evaluation of
$$
\probc{ (\mu_{1,t} - \mu_{2,t})^2 > c }{\mathcal{D}}
$$
for some $c>0$, which can be used for comparing the survival experiences of the two samples. Furthermore, we extend such methodology for the restricted to $[0,t]$ $k$-th moment of the survival time (t-kMST) defined as
$$
\mu_{1,t}^{(k)} =  k \int_0^t 
e^{-\xi_1 (s) } s^{k-1} \d s, \quad 
\mu_{2,t}^{(k)} =  k\int_0^t 
e^{-\xi_2 (s) } s^{k-1} \d s
$$
so the density corresponding to the difference of higher order central  moments or moments between groups can be calculated. For example, we can 
approximate the posterior distribution of the difference of the variances of survival times $\sigma^2_{1,t}-\sigma^2_{2,t}$, 
restricted to $[0,t]$, 
and use this to evaluate 
$$
\probc{ | \sigma^2_{1,t}  - \sigma^2_{2,t} | > c }{\mathcal{D}}. 
$$
for some $c>0$. 
\medskip

\section{Mixed moments of multivariate exponential functionals}\label{sec:mixed_moments}

Since our goal is to address a two-sample problem through a comparison of $\mu_{1,t}$ and $\mu_{2,t}$, we need to introduce a procedure for determining an approximation of their marginal and joint distributions conditionally on data that are subject to a right-censoring mechanism. A possible approximation technique arises from the knowledge of the marginal posterior moments of $\mu_{1,t}$ and $\mu_{2,t}$ and mixed posterior moments of $(\mu_{1,t},\mu_{2,t})$ up to a certain order. Since neutral to the right priors are conjugate also in the multivariate case  it is enough to examine the problem \textit{a priori}, see \cite{riva2018a} or supplementary material for a full review on NTR modeling. For our purposes it is convenient to introduce the $k$-th moment exponential functional of an increasing additive process $\xi_i$ restricted to $(s,t)$, for any $0\le s<t\le +\infty$, which is defined as 
\begin{equation}
	\label{eq:s_t_kth_expfunct}
	\mathcal{I}_{s,t}(\xi_i;k) := \mu_{i,t}^{(k)}-\mu_{i,s}^{(k)}
	=k \int_s^t e^{-\xi_i(u)} u^{k-1} \:\ddr u	.
\end{equation}
For simplicity, we will further set the notation $\mathcal{I}_{s,t}^r(\xi,k)=\left( \mathcal{I}_{s,t}(\xi,k) \right)^r$, $\mathcal{I}_{t}(\xi;k):=\mathcal{I}_{0,t}(\xi;k)$ and $\bar{\mathcal{I}}_{t}(\xi;k):=\mathcal{I}_{t,\infty}(\xi;k)$. The corresponding moments are denoted as
\begin{equation}
	\label{eq:moments_def}
	M_{s,t}^{(r)}(\xi_i;k) = \esp{ \mathcal{I}^r_{s,t}(\xi_i;k) },
\end{equation}
and analogously set 
$M_{t}^{(r)}(\xi_i;k)=M_{0,t}^{(r)}(\xi_i;k)$ and $\bar{M}_{t}^{(r)}(\xi;k)=M_{t,\infty}^{(r)}(\xi;k)$. The case with $k=1$ and any $r\ge 1$ has been investigated in \cite{salminen}, who prove the following recursive relationship
\begin{equation}
	\label{eq:salminenrecur}
	M_{s,t}^{(r)}(\xi_i;1) = r\int_s^t M_{u,t}^{(r-1)}(\xi_1;1)\:
	\edr^{-\{\psi_u(r) -\psi_u(r-1)\}}\,\d u. 
\end{equation}
A special case of \eqref{eq:salminenrecur} was originally established by \cite{epilijoiprunst}: it boils down to 
\begin{align*}\label{PerpMoment}
	M_\infty^{(r)}(\xi_i;1) &=
	r!\:\mathbb{E}\left[
	\int_0^\infty \cdots \int_{t_{r-1}}^\infty
	\prod_{j=1}^r \edr^{-(r+1-j)\{\xi(t_j)-\xi(t_{j-1})\}}
	\d t_r \cdots \d t_1
	\right]
	\\ & = r! \,
	\int_0^\infty \cdots \int_{t_{r-1}}^\infty
	\prod_{j=1}^r \edr^{- \{\psi_{t_j}(r+1-j) -\psi_{t_j}(r-j) \} }
	\d t_r \cdots \d t_1.
\end{align*}
and was used for evaluating an approximation of the density function of the random mean $\mu_{i,\infty}$ in a NTR model with a single sample or group, namely $d=1$. 

The $k$-th moment exponential functional in \eqref{eq:s_t_kth_expfunct} can be further extended to mixed moments related to a multivariate vectors of increasing additive processes $\bm{\xi}=(\xi_1,\ldots,\xi_d)$. These are crucial in the approximation procedure we propose as they are suited to capture properties of the model for survival times, with a multivariate neutral to the right prior. For the sake of keeping the notation simple, we let $\bm{k}=(k_1,\ldots,k_d)$ and $\bm{r}=(r_1,\ldots,r_d)$ vectors of non--negative integers. Hence, we can give the following
\smallskip

\begin{definition}
	For any $0\le s<t\le +\infty$ and $d>1$, the $\pmb{k}$--mixed moment exponential functional of $\bm{\xi}$ restricted to $(s,t)$ is 
	\begin{equation}
		\label{eq:k_mix_moment}
		M_{s,t}^{(\pmb{r})}(\pmb{\xi};\bm{k})=\mathbb{E}[\mathcal{I}_{s,t}^{r_1}(\xi_1;k_1)\:\cdots\: \mathcal{I}_{s,t}^{r_d}(\xi_d;k_d)]
	\end{equation}	
\end{definition}
For any vector $\bm{x}=(x_1,\ldots,x_d)$ and $i=1,\ldots,d$,  we introduce the following notation
\[
M_{s,t}^{(\bm{0})}(\bm{\xi};\bm{k})=1,\qquad M_{s,t}^{(\bm{r}_i^0)}(\bm{\xi};\bm{k})=M_{s,t}^{(\bm{r}_{-i})}
(\bm{\xi}_{-i};\bm{k}_{-i})
\] 
where  $\bm{x}_{-i}=(x_1,\ldots,x_{i-1},x_{i+1},\ldots,x_d)$ and $\bm{x}_{i}^0=(x_1,\ldots,x_{i-1},0,x_{i+1},\ldots,x_d)$. 
Similarly to $k$--th moment exponential functionals, we set 
$M_{t}^{(\pmb{r})}(\pmb{\xi};\bm{k})=M_{0,t}^{(\pmb{r})}(\pmb{\xi};\bm{k})$ and $\bar{M}_{t}^{(\pmb{r})}(\xi;\pmb{k})=M_{t,\infty}^{(\pmb{r})}(\pmb{\xi};\pmb{k})$. Such notation is useful for the implementation of the following recursion result which allows for the evaluation of $M_{s,t}^{(\pmb{r})}(\pmb{\xi};\pmb{k})$.
\smallskip

\begin{theorem}\label{teo1}
	If for any $\bm{r}$ such that $\min\{r_1,\ldots,r_d\}>0$ one has $\psi_t(\pmb{r})<\infty$ for any $t>0$, then the following holds true
	\begin{equation}
		\label{eq:recursion_mixed_moments}
		M_{s,t}^{(\pmb{r})}(\bm{\xi};\pmb{k})=
		\sum_{i=1}^d  r_i k_i \,
		\int_s^t M_{u,t}^{(\pmb{r}-\pmb{e}_i)}( \pmb{\xi};\pmb{k})\,
		\edr^{-\{ \psi_u(\pmb{r}) - \psi_u(\pmb{r}-\pmb{e}_i)\} }\,u^{k_i-1}\:
		\d u,
	\end{equation}
	where $\bm{e}_i$ is the vector with all zero entries but the $i$--th, which equals $1$.
\end{theorem}
\smallskip

\noindent 
If for some $i$ one has $r_i=0$, we have an analogous recursion with the $i$--th term vanishing. 
On the other hand we can take $ \xi_i \stackrel{\text{a.s.}}{=}\xi_l$ for $i\neq l$ and the above recursion is still valid. For example, if $\pmb{\xi}=(\xi_1,\xi_1)$ then the associated multivariate Laplace exponent is $\psi_u(r_1,r_2)=\psi_u(r_1+r_2)$ where $\psi_u$ is the Laplace exponent of $\xi_1$. A  similar notation can be used when a subset of entries of $\pmb{\xi}$ are almost surely equal.
\\
We remark that the above recursion generalizes the results of \cite{epilijoiprunst} and \cite{salminen} in two directions; first it correspond to functionals of multivariate vectors of subordinators and secondly goes beyond the exponential integral case due to the polynomial factors inside of the integrals when $k_i>1$, which allow for calculation of higher order moments for NTR priors.

\noindent As anticipated, the recursion in \eqref{eq:recursion_mixed_moments} holds true also \textit{a posteriori} and one just needs to replace the Laplace transform, say 
$\edr^{-\left( \psi_u(\pmb{r}) - \psi_u(\pmb{\ell}) \right)}$, with the corresponding to the posterior representation of multivariate neutral to the right priors for partially exchangeable survival times. Note that such a transform includes also the jumps occurring at the exact (uncensored) observations. For example, if we confine ourselves to the case $d=2$ we let $\pmb{\mathcal{D}}=\{(\Delta_{i,j},t_{i,j}):\:i=1,\ldots,n_j;\: j=1,2\}$ denote the data, we arrange the $T_{i,j}$ in increasing order and associate them the corresponding $\Delta_{i,j}$. Additionally, $O(t)$ is the set of observations in $\pmb{\mathcal{D}}$ before $t$, $E_i(t)$ the set of exact observations in sample $i$ of $\pmb{\mathcal{D}}$ before $t$ and $\tilde{n}_{i,\tilde{t}},\tilde{m}_{i,\tilde{t}}$ constants depending on the data $\pmb{\mathcal{D}}$, $i\in\left\{1,2\right\}$, $0<\tilde{t}$. Hence, the Laplace transform that is involved in the recursion \eqref{eq:recursion_mixed_moments} has the following simple form
\begin{align*}
	\edr^{-\{\psi_u(\bm{r}|\pmb{\mathcal{D}})-\psi_u(\bm{\ell}|\pmb{\mathcal{D}})\}}
	&=
	\edr^{\sum_{\tilde{t}\in O(t)}\{ \psi_{\tilde{t}}\left( r_1+\tilde{n}_{1,\tilde{t}},r_2+\tilde{n}_{2,\tilde{t}} \right) - \psi_{\tilde{t}}\left( \ell_1+\tilde{n}_{1,\tilde{t}},\ell_2+\tilde{n}_{2,\tilde{t}} \right)  \}}
	\\
	& \quad \times\,\prod_{i=1}^2 
	\prod_{ \tilde{t}\in E_i(t) }\frac{ \psi_{\tilde{t}}(r_1+\tilde{m}_{1,\tilde{t}}+1,
		r_2+\tilde{m}_{2,\tilde{t}}) - \psi_{\tilde{t}}
		( r_1+\tilde{m}_{1,\tilde{t}},2_2+\tilde{m}_{2,\tilde{t}}) }{ \psi_{\tilde{t}}(\ell_1+\tilde{m}_{1,\tilde{t}}+1,\ell_2+\tilde{m}_{2,\tilde{t}}) - \psi_{\tilde{t}}(\ell_1+\tilde{m}_{1,\tilde{t}},\ell_2+\tilde{m}_{2,\tilde{t}})  }
\end{align*}
In particular, the Laplace exponents in the above expression can be evaluated for vectors of compound random measures through a Monte Carlo procedure whenever the Laplace exponent of the directing L\'evy measure is available at positive integers. This is the case of the Log-Beta process. Even more, simple choices of the score distribution might result in no need for the Monte-Carlo approximation. Such an example is the $(\pmb{\pi}_1,\pmb{\pi}_2)$--stratified-compound Log-Beta process of section \ref{sec:mixed_moments}, which has categorical score distributions. 

\begin{center}
\begin{figure}[h!]
\begin{center}
\begin{tabular}{cc}
\includegraphics[width=0.49\linewidth]{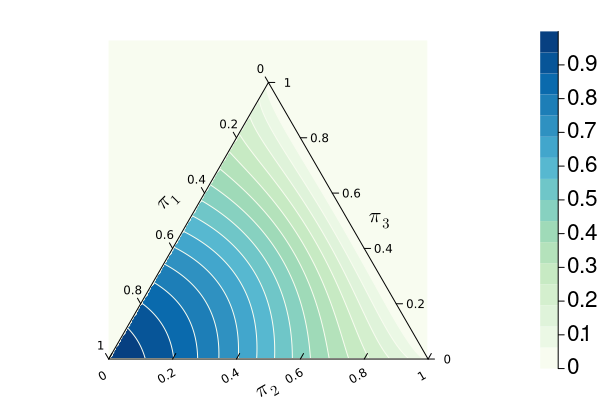} &
\includegraphics[width=0.49\linewidth]{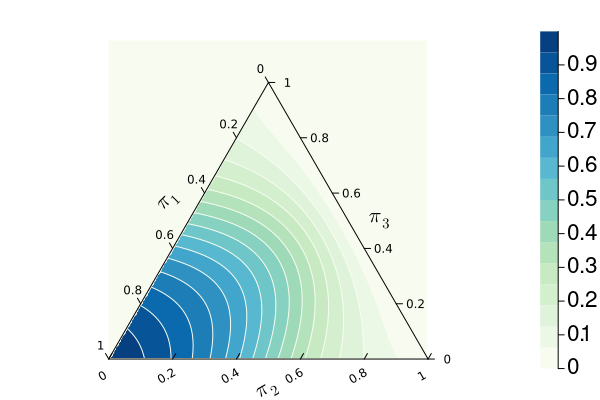} \\
(i) & (ii)
\end{tabular}
\caption{Correlation between RMSTs for (i) $t=5$ and (ii) $t=10$ in $\bm{\pi}$--compound Beta-Stacy model.} 
\label{meancorrPlot}
\end{center}
\end{figure}
\end{center}

A simple application of Theorem \ref{teo1} leads to a closed form expression of the correlation between the restricted random means, namely 
\begin{align*}
	&\text{Corr}\left(\mu_{1,t},\mu_{2,t}\right) = \frac{  M_t^{(1,1)}(\pmb{\xi};\pmb{1}) - M_t^{(1)}(\xi_1;1)M_t^{(1)}(\xi_2;1) }{
		\sqrt{ \left( M_t^{(2)}(\xi_1;1) - M_t^{(1)}(\xi_1;1)^2  \right) \left( M_t^{(2)}(\xi_2;1) - M_t^{(1)}(\xi_2;1)^2  \right)  
	}}
\end{align*}
We show in Figure \ref{meancorrPlot} such correlation  for the $\pmb{\pi}$--compound Beta-Stacy process for different values of $\pmb{\pi}$. Furthermore the correlation between random variances $\sigma^2_{i,t}=\mu_{i,t}^{(2)}-\mu_{i,t}^2$ can be evaluated, as detailed in Section~\ref{AppD} of the Appendix.

If one moves on to considering restricted survival times, 
$\bm{k}$--mixed moments easily follow from previous calculations. 
It suffices to consider linear combinations of powers of the $k$-th moments $\mu_{i,t}^{(k)}=k\int_0^t e^{-\xi_i(u) }u^{k-1} \d u$ arising from a vector of subordinators with entries $\xi_i$, $1\leq i\leq d$ that can possibly be repeated, i.e. $\xi_i=\xi_j$ for $i\neq j$. This will be helpful for the evaluation of moments of difference of central moments such as $\mu_{1,t}-\mu_{2,t}$, which, as anticipated, are quite natural to consider when comparing two samples.

Let us finally agree on some notation. For any two vectors $\bm{\ell}$ and $\bm{k}$ in $\N^d$, with $(\bm{\ell k})_m$ we denote the vector with entries in positions $i_1,\ldots,i_m$ equal $\ell_{i_1} k_{j_1},\ldots, \ell_{i_m} k_{j_m}$, respectively, with $m\le d$. Moreover, if $\bm{\xi}=(\xi_1,\ldots,\xi_d)$ is a vector of subordinators, for any $i_1,\ldots,i_m\in\{1,\ldots,d\}$ we let $\bm{\xi}_i=(\xi_{i_1},\ldots,\xi_{i_m})$. 
\smallskip

\begin{proposition}\label{prop2}
Let $t>0$, $d,m,k_1,r_1,\cdots,k_m,r_m\in \mathbb{N}\setminus\{0\}$, $\pmb{\xi}$ a $d$-variate vector of subordinators, $i_1,\ldots,i_m \in \left\{1,\ldots,d\right\}$, and $a_1,\ldots,a_m\in\re$. Then, in the $\text{NTR}(\pmb{\xi})$  model, if $X_l = \left( \mu_{i_l,t}^{(k_l)} \right)^{r_l}$ we have that 
\begin{align*}
\esp{ \Big(\sum_{l=1}^m a_{l} X_l \Big)^n} =\sum_{\bm{\ell}\in\mathcal{S}_{m,n}} 
		\binom{n}{\ell_1\:\cdots\:\ell_m} 
a_{1}^{\ell_1}\,\cdots\,a_{m}^{\ell_m} \: 
M_t^{(\bm{\ell r})}(\pmb{\xi}_i;\pmb{k})
\end{align*}
where $\mathcal{S}_{m,n}:=\{\bm{\ell}\in\N^m:\: \ell_1+\,\cdots\,+\ell_m=n\}$. 
\end{proposition}
\smallskip

\noindent As anticipated, a useful application of the above result is when $d=2$, $(i_1,i_2)=(1,2)$ and $a_{1}=-a_{2}=1$, as it yields the following
\smallskip

\begin{corollary}\label{cor1}
	Let $n\in \mathbb{N}$. Then 
	\begin{align*}
		\esp{ (\mu_{1,t}-\mu_{2,t})^n } &=  \sum_{\ell=0}^n (-1)^{\ell}  \binom{n}{\ell} M_t^{(1,1)}(\bm{\xi};\bm{k}_\ell)
	\end{align*}
	where $\bm{k}_\ell=(n-\ell,\ell)$. 
\end{corollary}
Furthermore, we can obtain an analogous result for the variances
\begin{corollary}\label{cor2}
	Let $n\in \mathbb{N}$. Then 
	\begin{align*}
		\esp{ (\sigma^2_{1,t}-\sigma^2_{2,t})^n } &= \esp{ (\mu^{(2)}_{1,t} -\mu^2_{1,t}-\mu^{(2)}_{2,t} + \mu^2_{2,t} )^n } \\ &=  \sum_{\bm{\ell}\in\mathcal{S}_{m,n}} (-1)^{\ell_2+\ell_3}  \binom{n}{\ell_1,\ell_2,\ell_3,\ell_4} M_t^{(2,1,2,1)}\left(\bm{\xi};(\ell_1,2\ell_2,\ell_3,2\ell_4)\right)
	\end{align*}
\end{corollary}
\smallskip


It is worth noting that the restricted survival times 
correspond to random variables whose distribution is not necessarily absolutely continuous with respect to Lebesgue measure; for instance, this is the case when $\nu_{t_0}$ has a point mass different from $\pmb{0}$ at some time point $t_0>0$. Such a condition is met by posterior subordinators, given exact observations, as shown in Theorem~\ref{posteo} of Section~\ref{AppB} in the Appendix. On the other hand, moments in Proposition~\ref{prop2} can be used to perform density estimation when the distribution of the corresponding random variable is absolutely continuous with respect to Lebesgue measure. This motivates the interest in the following proposition, which gives sufficient conditions for absolute continuity, with respect to the Lebesgue measure, of the probability distribution of $\sum_{l=1}^m a_l \{\mu_{i_l,t}^{(k_l)}\}^{r_l}$. 
\smallskip

\begin{proposition}\label{prop4}
Let $t>0$, $d\in \mathbb{N}\setminus \{0\}$, $\pmb{\xi}$ a $d$-variate vector of subordinators with intensity $\nu$ such that for some $0\leq s<t$
	\begin{equation*}
		\int_{\re_+^d} \left( \| \pmb{x} \|^2 \wedge 1 \right) \left( \nu_t( \d \pmb{x}  ) -\nu_s( \d \pmb{x}  ) \right) < \infty,
\end{equation*}
then, in the $\text{NTR}(\pmb{\xi})$  model, r.v.'s of the form   $$ \sum_{l=1}^m a_l\left( \mu_{i_l,t}^{(k_l)} \right)^{r_l}$$ 
are absolutely continuous with respect to Lebesgue measure in $\re$, where $m,k_1,r_1,\cdots,k_m,r_m\in \mathbb{N}\setminus\{0\}$, $i_1,\ldots,i_m \in \left\{1,\ldots,d\right\}$ and $a_1,\ldots,a_m\in\re$.
\end{proposition}
\smallskip


\section{Data studies}

We focus on density estimation for the mean and variance difference across two populations. With such aim, we use the recursion formulas for multivariate k-moment exponential integrals of Proposition \ref{teo1} to calculate the moments of such difference by use of Corollary \ref{cor1}, denote the moments of interest $\left\{c^{(k)}\right\}_{i=1}^\infty$. Following such calculation we perform a discrete maximum entropy approximation which solves the problem of optimising Shannon entropy of a discrete random variable $X$ taking values on a mesh $\pmb{x} = (x_1,\ldots,x_m) \in \mathbb{R}^m$, $x_i<x_{i+1}$, $1\leq i \leq m\in\mathbb{N}$, with corresponding probabilities $p_X(x_i)=p_i=\mathbb{P}\left[X=x_i \right],\,1\leq i \leq m$. We denote the Shannon entropy of such $X$ as  

$$
H_{\pmb{x}}(\pmb{p}) = 
-\sum_{x\in\pmb{x}} p_X(x)\log\left( p_X(x) \right)  =
-\sum_{i=1}^m p_i\log\left( p_i \right) 
$$

\noindent
subject to $N$ moment constraints 
\begin{equation}\label{momConstr}
\mathbb{E}\left[X^k\right]=
c^{(k)},\quad 1\leq  k \leq N,
\end{equation}  
see \cite{jaynes}. Such optimization task for finding the probabilities $\left\{ p_X(x),\, x\in\pmb{x}\right\}$ is well understood, \cite{boyd} gives a full treatment of the topic, and software for solving it is readily available, see \url{https://web.stanford.edu/~boyd/software.html} for a list. In Algorithm \ref{algo} we present pseudocode for the density estimations procedure for $\mu_{1,t}-\mu_{2,t}$.
\\

\begin{algorithm}
\KwIn{ Multiple-sample $\text{NTR}(\pmb{\xi})$ model, mesh of values $\pmb{x}\subset \mathrm{R}$ and integer $N$ for the number of moment constraints to consider.}
\KwOut{ Density approximation for $\mu_{1,t}-\mu_{2,t}$ .}
Using the recursion in Theorem \ref{teo1} calculate $M_t^{(n-l,l)}\left( \pmb{\xi}; \pmb{1} \right)$, $l\in\left\{0,\ldots,n\right\}$.
\\
Calculate $c^{(k)}=\esp{ \left( \mu_{1,t} - \mu_{2,t} \right)^k  }$ for $k\in \left\{1,\ldots, n\right\}$ using Proposition \ref{prop2}.
\\
Let $\pmb{p}^\star$ be the maximum of $H_{\pmb{x}}(\pmb{p})$ subject to the moment constraints $\mathbb{E}\left[X^k\right]=c^{(k)}$. \\
Approximate the density with $\hat{f}(x)\approx p^\star_j (x_j-x_{j-1})$ for $x\in(x_{j-1},x_j]$.
\caption{Density estimation for moment difference between populations}
\label{algo}
\end{algorithm}

Further comparison of the survival experience based on density approximations for differences of moments related to variance, skewness or kurtosis can be performed in a similar approach. For instance in the case of the difference of variances across populations and $\sigma_{1,t}^2-\sigma_{2,t}^2$, step 1 of Algorithm \ref{algo} requires calculation of $ M_t^{(l_1,2l_2,l_3,2l_4)}\left( \left( \xi_1,\xi_2,\xi_1,\xi_2 \right); \left(2,1,2,1 \right) \right),\; \pmb{l}\in \left\{ \left\{0,\ldots,n\right\}^4 \,:\, l_1+l_2+l_3+l_4=n\right\}$ and in step 2 setting $ c^{k}=\esp{ \left( \sigma_{1,t}^2 - \sigma_{2,t}^2 \right)^k  }$. 

A simulated and real data study are presented next.

\subsection{Simulation study}\label{SimStudy}

We consider two multiple samples with equal number $n\in\mathbb{N}$ of observations each. Observations are drawn drawn from mixtures of Weibull distributions. Let $f_{W(c_1,c_2)}(\cdot)$ be a Weibull density with shape $c_1$ and scale $c_2$. We generate
\begin{equation*}
\left\{ Y_{1,i} \right\}_{i=1}^n \stackrel{\text{i.i.d.}}{\sim} \mathcal{L}(f_1) \; \perp \; \left\{ Y_{2,i} \right\}_{i=1}^n \stackrel{\text{i.i.d.}}{\sim} \mathcal{L}(f_2) 
\end{equation*}
with
\begin{align*}
f_1(\cdot)=\frac{1}{2}\left( f_{W(2.1,5)}(\cdot)+ f_{W(1.2,5.5)}(\cdot) \right),\quad 
f_2(\cdot)= \frac{1}{2}\left( f_{W(2.1,5)}(\cdot)+f_{W(5.3,4.75)}(\cdot) \right).
\end{align*}
Furthermore we simulate censoring times 
\begin{equation*}
\left\{ C_{1,i} \right\}_{i=1}^n \stackrel{\text{i.i.d.}}{\sim} \text{Exponential}(\theta_1) \; \perp \; \left\{ Y_{2,i} \right\}_{i=1}^n \stackrel{\text{i.i.d.}}{\sim} \text{Exponential}(\theta_2)
\end{equation*}
where $\theta_1,\theta_2>0$ were fixed via the Robbins-Monro algorithm \citep{RobbinsMonro} targeting a probability of $0.8$ for censoring, i.e. $Y_{k,i} < C_{k,i}$, $k\in\left\{1,2\right\}$; the algorithm was run over 10000 iterations with initial value 3 and step sizes $i^{0.75}$ for iteration $i\in\mathbb{N}$.

We fitted the multiple-sample NTR $(\pmb{\pi}_1,\pmb{\pi}_2)$--stratified-compound Beta-Stacy estimators for the survival functions. We chose the hyperparameter $\beta$, which corresponds to the \textit{a priori} centering for the NTR distribution, to be the survival function of an $\text{Exponential}(0.3)$ and $\gamma=1$; the latter reflects a moderate \textit{a priori} variance for the survival, more precisely  variance $\beta(t)(1-\beta(t))/(1+\gamma)$ at time $t$. The parameters $\pmb{\pi}_1,$ $\pmb{\pi}_2$ and threshold 
$\tau$ were fixed via maximum a posteriori using the marginal likelihood with the vector of subordinators integrated out, see Section~\ref{AppB} of the Appendix for details.

\begin{center}
\begin{figure}[h!]
\begin{center}
\begin{tabular}{cc}
\includegraphics[width=0.49\linewidth]{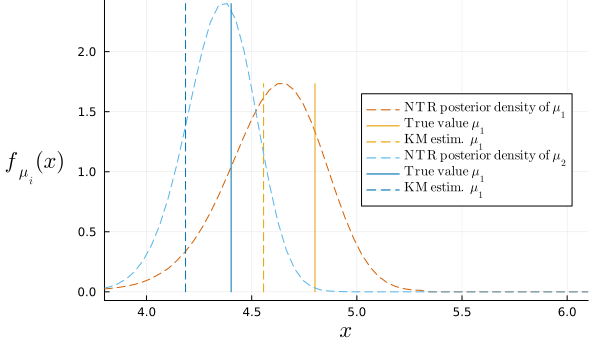} &
\includegraphics[width=0.49\linewidth]{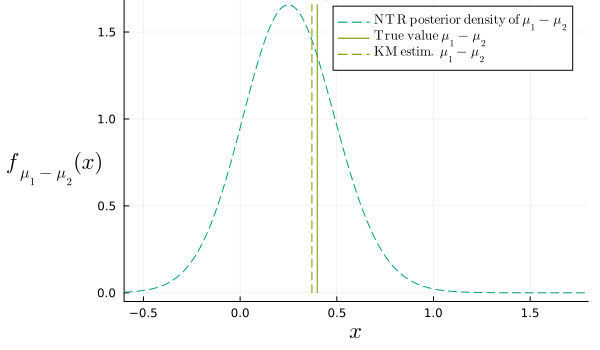} \\
(i) & (iii)
\end{tabular}
\\
\begin{tabular}{cc}
\includegraphics[width=0.49\linewidth]{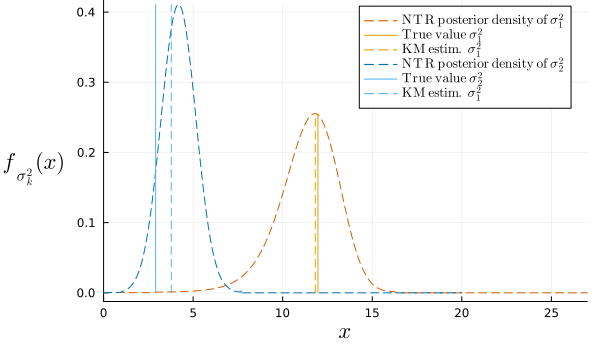} &
\includegraphics[width=0.49\linewidth]{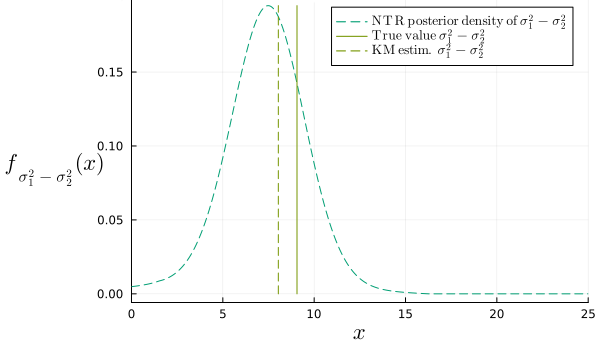} \\
(ii) & (iv)
\end{tabular}
\caption{For the simulation study described in Section \ref{SimStudy}, with $n=300$ observations we show maximum entropy density estimates of: the  marginal distribution of (i) restricted means and (ii) variances across populations and the difference between of restricted (iii) means and (iv) variances.}\label{sim300study}
\end{center}
\end{figure}
\end{center}

\begin{center}
\begin{figure}[h!]
\begin{center}
\begin{tabular}{cc}
\includegraphics[width=0.49\linewidth]{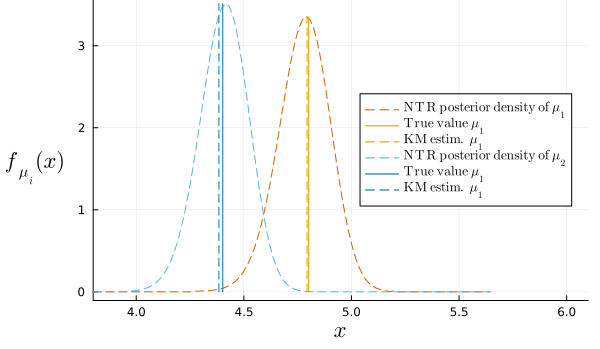} &
\includegraphics[width=0.49\linewidth]{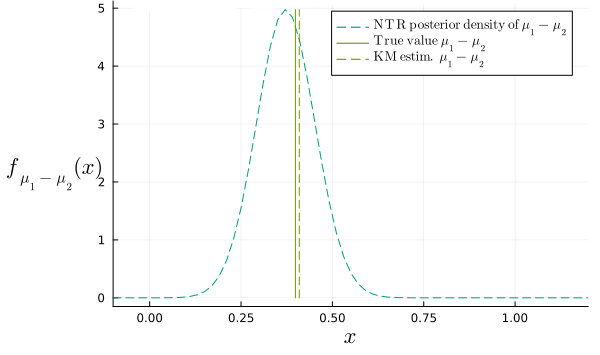} \\
(i) & (iii)
\end{tabular}
\\
\begin{tabular}{cc}
\includegraphics[width=0.49\linewidth]{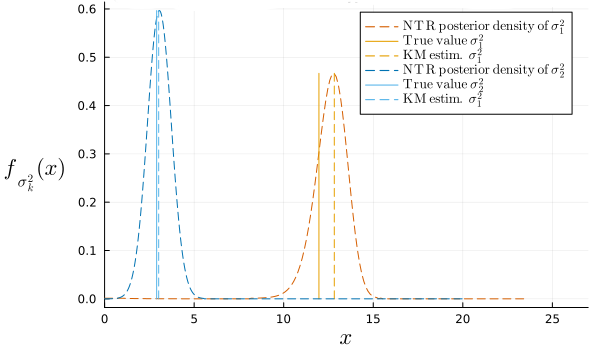} &
\includegraphics[width=0.49\linewidth]{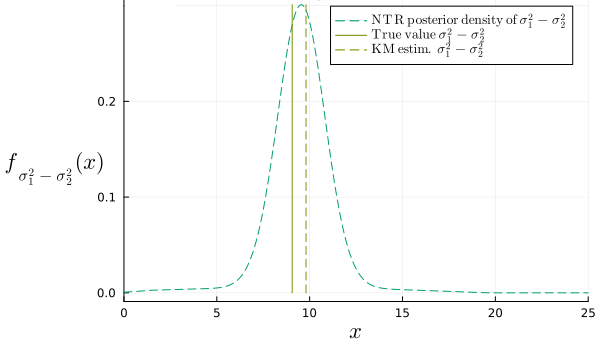} \\
(ii) & (iv)
\end{tabular}
\caption{For the simulation study described in Section \ref{SimStudy}, with $n=3000$ observations we show maximum entropy density estimates of: the  marginal distribution of (i) restricted means and (ii) variances across populations and the difference between of restricted (iii) means and (iv) variances.}\label{sim3000study}
\end{center}
\end{figure}
\end{center}

The simulation studies were performed for $n\in\left\{300,3000\right\}$ observations in each sample population. We show in Figures \ref{sim300study} and \ref{sim3000study} posterior restricted mean and variance density estimates where the cutoff restriction in time was $t_f=30$, a value well above the $99\%$ quantiles $17.14$ and $9.14$ of respectively $\mathcal{L}(f_1)$ and $\mathcal{L}(f_2)$. We can see in the figures density fits for the marginal quantities as well as their difference using Algorithm \ref{algo} with $6$ moment constraints, a higher number of constraints resulted in negligible changes of the density. For computation we used a mesh $\pmb{x}$  which was suitably chosen for each density based on the first two moments and Chebyshev inequality, for example $600$ equidistant points between $-6$ and $6$ for the difference of means and between $-10$ and $30$ for the variance.

\begin{center}
\begin{figure}[h!]
\begin{center}
\includegraphics[width=0.49\linewidth]{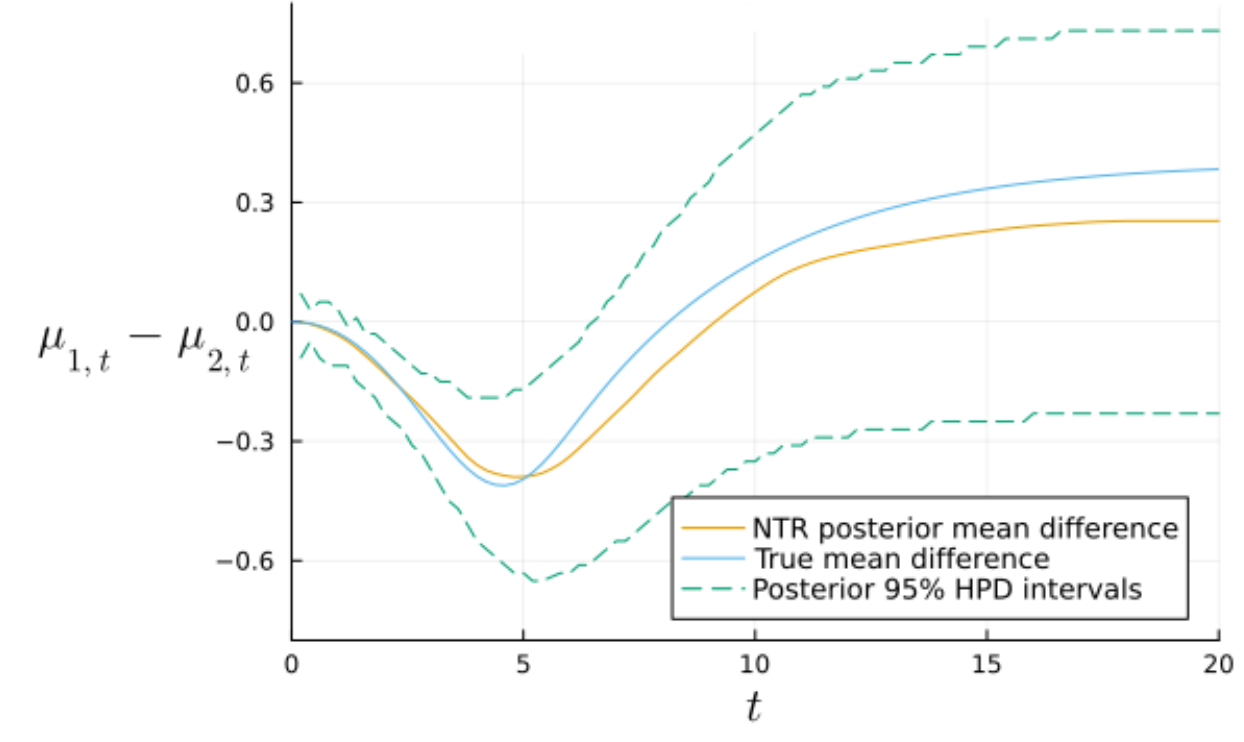}
\includegraphics[width=0.49\linewidth]{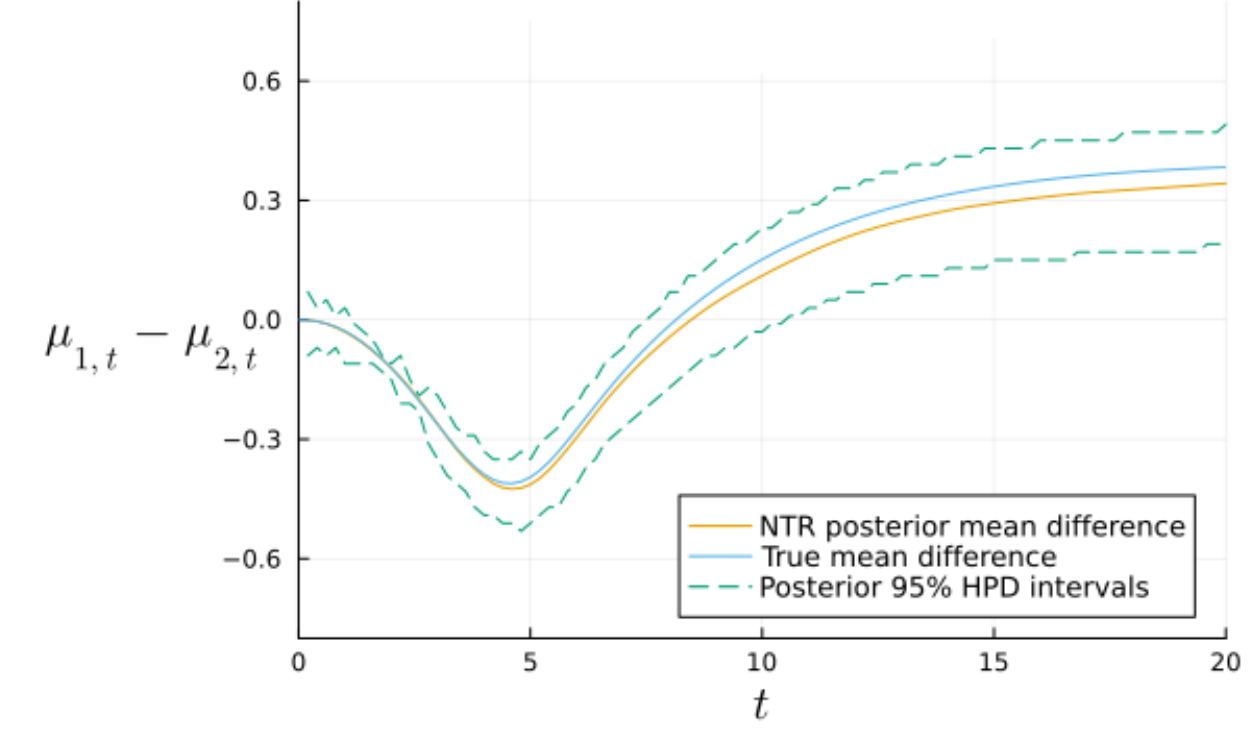}
\caption{For the simulation study described in Section \ref{SimStudy}, we show pointwise highest posterior density intervals for the differences of RMST's with (i) $n=300$ and $n=3000$ observations. }
\label{hpdPlots}
\end{center}
\end{figure}
\end{center}

For a general time $t>0$, we show in Figure \ref{hpdPlots}  $95\%$ highest posterior density regions, see \cite{hyndman} where calculation procedures are discussed, for the difference of RMST's $\mu_{1,t}-\mu_{2,t}$. We observe that the multiple sample NTR model with the density estimation procedure does a good work on quantifying uncertainty as the number of observation increases from $300$ to $3000$.

\newpage

\subsection{AIDS real data study}\label{aidsec}

\begin{center}
\begin{figure}[h!]
\begin{center}
\includegraphics[width=0.9\linewidth]{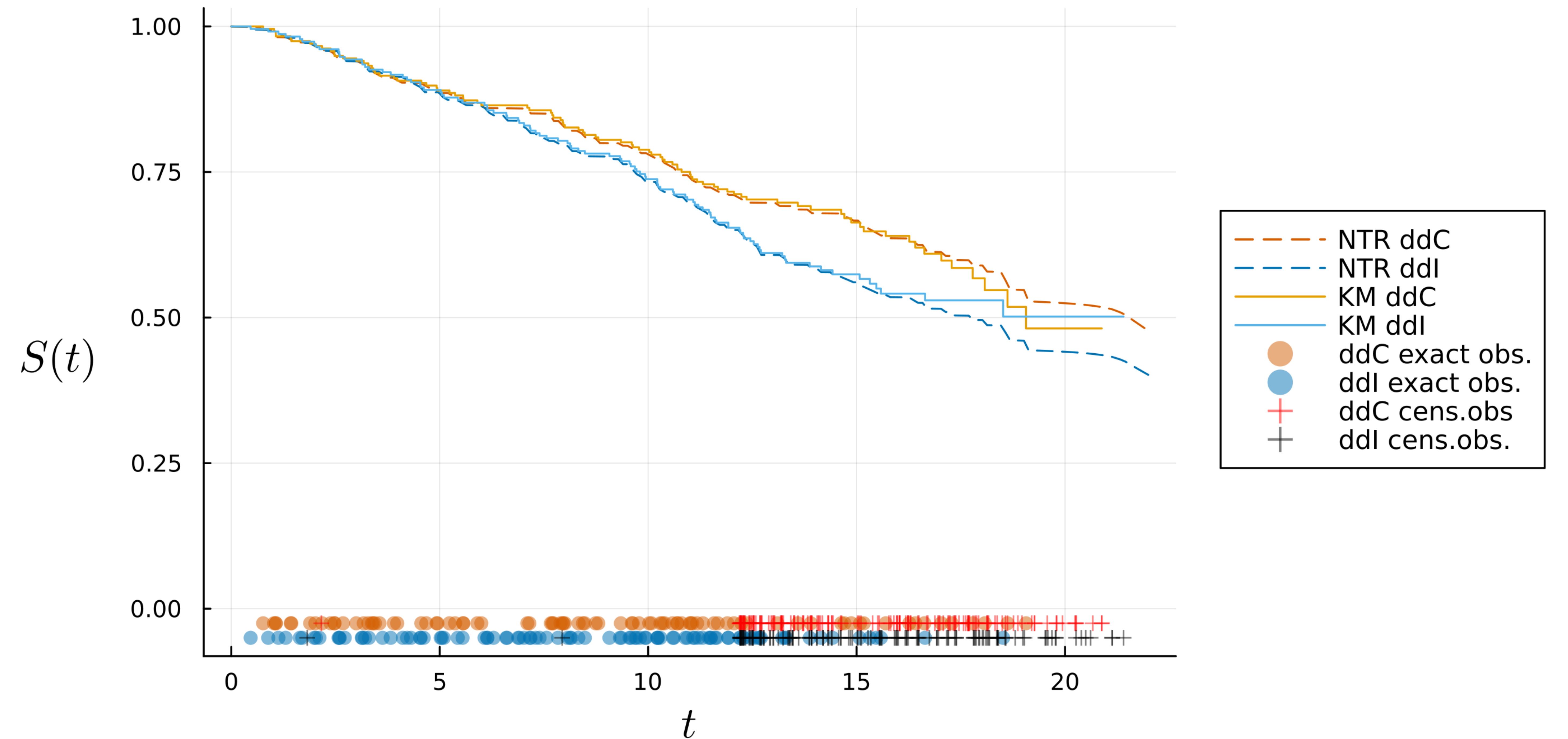}
\caption{Neutral to the right and Kaplan-Meier survival curve estimates for right censored observations of AIDS patients treated with ddC and ddI drug treatments.}\label{aidsSurvfit}
\end{center}
\end{figure}
\end{center}

\begin{center}
\begin{figure}[h!]
\begin{center}
\includegraphics[width=0.9\linewidth]{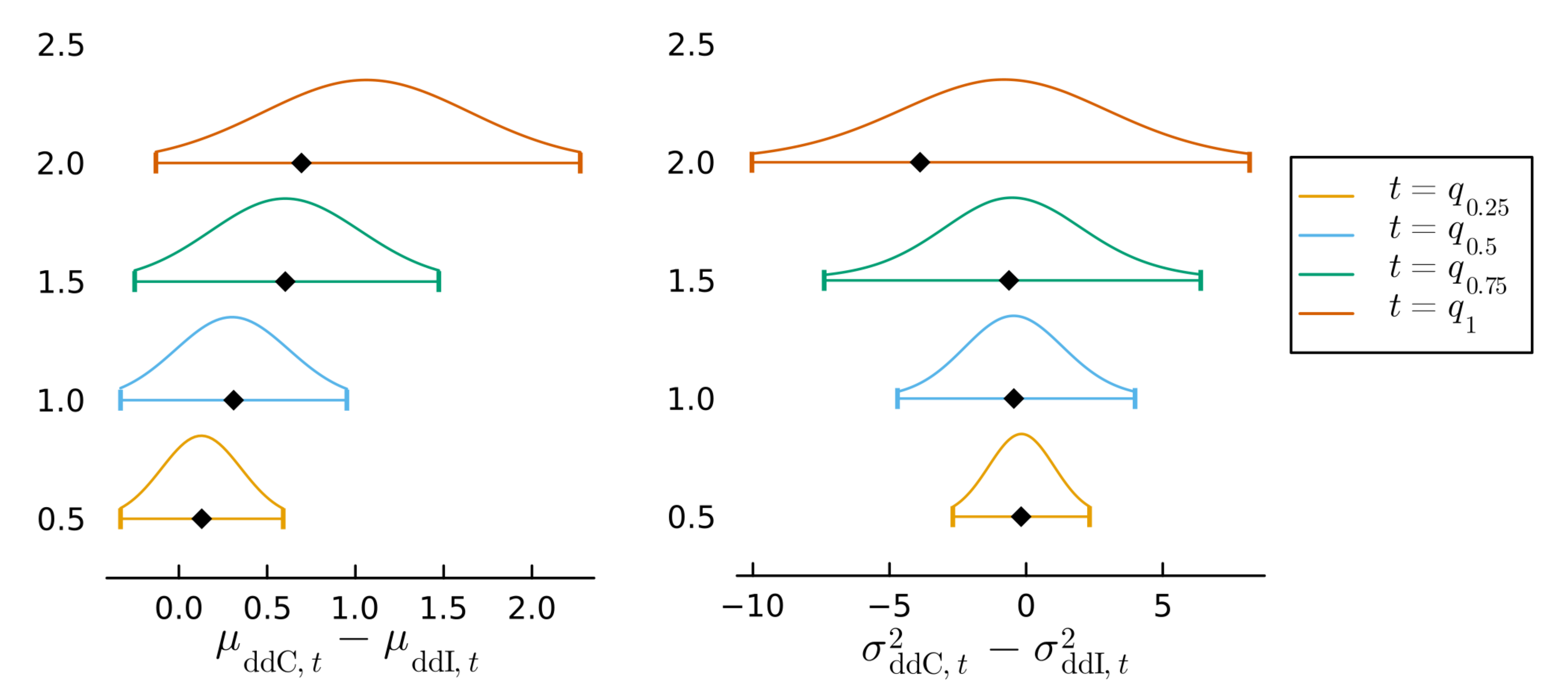}
\caption{95\% highest posterior density intervals (solid lines) and point estimates obtained from marginal Kaplan-Meier fits (black diamond markers) for mean and variance differences restricted to $[0,t]$ with $t\in\left\{q_{0.25},q_{0.5},q_{0.75},q_1\right\}$.}\label{aidsHPDintervs}
\end{center}
\end{figure}
\end{center}

We consider the randomized clinical trial in \cite{aids} where patients were subject to two antiretroviral drugs, didanosine (ddI) and zalcitabine (ddC). The data is available in the "JM" R programming language package  \citep{jm}. Figure \ref{aidsSurvfit} shows marginal Kaplan-Meier and multiple-sample NTR $(\pmb{\pi}_1,\pmb{\pi}_2)$--stratified-compound Beta-Stacy estimators for the survival functions. We chose the a priori baseline survival, determined by the hyperparameter $\beta$, to be an $\text{Exponential}(0.1)$  distribution, and hyperparameter $\gamma=1$ which reflects a moderate a priori variance around the baseline, as in the previous simulation study. The $\pmb{\pi}_1,\,\pmb{\pi}_2$ and thresholding parameter $\tau$ were fixed via maximum a posteriori using the marginal likelihood, as in the simulation studies. We summarize inference in figure \ref{aidsHPDintervs} where we plot posterior densities over the 95\% highest posterior density (HPD) intervals, associated to moment differences $\mu_{1,t_r}-\mu_{2,t_r}$ and $\sigma_{1,t_r}^2-\sigma_{2,t_r}^2$ with $t_r\in\left\{ q_{0.25},q_{0.5},q_{0.75},q_1\right\}$, where $q_\alpha$ is the $\alpha\%$ quantile of the pooled data. Point estimates based on marginal Kaplan-Meier estimates are also shown for comparison. The posterior densities and HPD intervals were obtained numerically from the implementation of Algorithm \ref{algo} where the number of moments constraints was chosen incrementally until the difference in supremum norm in $[0,t_r]$ of the produced densities with $N$ and $N-1$ constraints was less than $0.1$; on the other hand, the mesh $\pmb{x}$ was chosen to consist of $600$ equidistant points between $-6$ and $6$ for difference of means density estimation and similarly $1200$ points between $-21$ and $21$ for the difference of variances. We observe that as more observations are considered, when increasing the time restriction, HPD's for $\mu_{1,t_r}-\mu_{2,t_r}$ lie in the positive real line, indicating a bigger mean survival time for patients with the ddC; on the other hand  HPD's for $\sigma_{1,t}^2-\sigma_{2,t}^2$ are centered around the origin indicating similar variances across populations.

%
%

\section{Discussion}

We introduced the use of exponential integral functionals for Bayesian non parametric applications in survival analysis based on NTR priors in three main directions: 1) We extended current frameworks onto a multivariate setting which allows for possibly dependent increasing additive process and motivates calculation of crossed moments of exponential integrals. From the BNP applications perspective this makes possible the calculation of posterior moments for multiple-sample. 2) We introduce $k$-th moment exponential integrals which relate to random moments of NTR models. 3) We considered the exponential integrals restricted to arbitrary interval so in particular calculation of restricted mean survival time in the NTR models can be studies. Recursion formulas for calculation of the novel $\pmb{k}$-mixed moment exponential functional were derived. Furthermore, in the multiple NTR model setting, we show the use of such quantities to perform posterior calculation of the correlation between populations, density estimation for restricted moments and their difference across samples and associated highest density regions. Flexibility and tractability of NTR priors show the usefulness of the proposed methodology in simulated and real data examples.

\clearpage
\newpage

\section*{Appendix}

\renewcommand{\thesection}{A}
\renewcommand{\thesubsection}{A.\arabic{subsection}}
\renewcommand{\theequation}{\Alph{subsection}.\arabic{equation}} 
\setcounter{equation}{0}

\section{Proofs of main results.}

\subsection{Proof of Proposition 1}\label{proof1}

\begin{proof} 
A simple change of variable implies that
$$\mathcal{I}_{s,t} (\xi_i,k)=
k \int_s^t  e^{-\xi_i(l)} l^{k-1} dl =
k e^{ -\xi_i(s)}\int_0^{t-s} e^{-\left(\xi_i (l+s)-\xi_i (s) \right)} (l+s)^{k-1} dl,
$$
$i=\left\{1,\ldots,n\right\}$, so by using the independent increment property of dependent subordinators

\begin{align*}
& M_{s,t}^{(\pmb{r})}(\bm{\xi};\pmb{k}) = \esp{ \mathcal{I}_{s,t}^{r_1} (\xi_1,k_1) \cdots \mathcal{I}_{s,t}^{r_n}(\xi_n,k_n) } =\mathbb{E}\left[ \left( k_1 \int_0^{t-s} e^{-\left(\xi_1 (l+s)-\xi_1 (s) \right)} (l+s)^{k_1-1} dl \right)^{r_1} \right.
\\ & \cdots \left. \left( k_n \int_0^{t-s} e^{-\left(\xi_n (l+s)-\xi_n (s) \right)} (l+s)^{k_n-1} dl \right)^{r_n} \right] e^{-\psi_s(\pmb{r})} \numberthis \label{indepincreeq}
\end{align*} 
On the other hand, for every $r_1,r_2,\ldots,r_n>0$, 
\begin{align*}
&\frac{d}{du} M_{u,t}^{(\pmb{r})}(\bm{\xi};\pmb{k}) = -r_1 k_1\esp{ \mathcal{I}_{u,t}^{r_1-1} (\xi_1,k_1)\mathcal{I}_{u,t}^{r_2}(\xi_2,k_2)\cdots \mathcal{I}_{u,t}^{r_n}(\xi_n,k_n)  e^{-\xi_1(u)}u^{k_1-1} }
\\ &  - \ldots - r_nk_n\esp{ \mathcal{I}_{u,t}^{(k_1)}(\xi_1)^{r_1}\mathcal{I}_{u,t}^{(k_2)}(\xi_2)^{r_2} \cdots \mathcal{I}_{u,t}^{(k_n)}(\xi_n)^{r_n-1}e^{-\xi_n(u)}u^{k_n-1}}
\end{align*}
We focus on the first term, as the rest can be treated analogously, and use the change of variable above so
\begin{align*}
& \esp{ \mathcal{I}_{u,t}^{(k_1)}(\xi_1)^{r_1-1}\mathcal{I}_{u,t}^{(k_2)}(\xi_2)^{r_2}\mathcal{I}_{u,t}^{(k_n)}(\xi_n)^{r_n} e^{-\xi_1(u)}u^{k_1-1} } =
\\& \mathbb{E}\left[ \left( k_1 \int_0^{t-u} e^{-\left( \xi_1(l+u)-\xi_1(u) \right) }(l+u)^{k_1-1} \d l\right)^{r_1-1} \right. \left( k_2\int_0^{t-u} e^{-\left( \xi_2(l+u)-\xi_2(u) \right) } (l+u)^{k_2-1} \d l \right)^{r_2}
\\ & \quad \; \cdots \left(k_n\int_0^{t-u} e^{-\left( \xi_n(l+u)-\xi_n(u) \right) } (l+u)^{k_n-1} \d l \right)^{r_n}  \Bigg] e^{-\psi_u(\pmb{r})}u^{k_1-1}
\\& 
\stackrel{\eqref{indepincreeq}}{=} M_{u,t}^{(\pmb{r}-\pmb{e}_1)}\left( \bm{\xi};\pmb{k} \right)e^{-\left( \psi_u(\pmb{r}) - \psi_u(\pmb{r}-\pmb{e}_1) \right) }u^{k_1-1}.
\end{align*}
It follows that $ \frac{d}{du}  M_{u,t}^{(\pmb{r})}(\bm{\xi};\pmb{k})  = -\sum_{i=1}^n r_i k_i M_{u,t}^{(\pmb{r}-\pmb{e}_1)}\left( \bm{\xi};\pmb{k} \right)e^{-\left( \psi_u(\pmb{r}) - \psi_u(\pmb{r}-\pmb{e}_i) \right) }u^{k_i-1}$ and integrating between $s$ and $t$ we conclude that $M_{s,t}^{(\pmb{r})}(\bm{\xi};\pmb{k})=
\sum_{i=1}^n  r_i k_i  \int_s^t M_{u,t}^{(\pmb{r}-\pmb{e}_1)}\left( \bm{\xi};\pmb{k} \right) e^{-\left( \psi_u(\pmb{r}) - \psi_u(\pmb{r}-\pmb{e}_i) \right) }u^{k_i-1}
 \d u$.
\end{proof}

\subsection{Proof of Proposition 1}\label{proof2}

\begin{proof}
\begin{align*}
\espc{ \left( \sum_{\ell=1}^m a_{\ell} X_\ell \right)^n}{\pmb{\xi}} &= \sum_{\bm{\ell}\in\mathcal{S}_{m,n}} \binom{n}{\ell_1\:\cdots\:\ell_m} a_{1}^{\ell_1}\,\cdots\,a_{m}^{\ell_m} \: \espc{
(\mu_{i_1,t}^{(k_1)})^{l_1 r_1}\cdots (\mu_{i_m,t}^{(k_m)})^{l_m r_m}
}{\pmb{\xi}}
\\ & =  \sum_{\bm{\ell}\in\mathcal{S}_{m,n}} \binom{n}{\ell_1\:\cdots\:\ell_m} a_{1}^{\ell_1}\,\cdots\,a_{m}^{\ell_m} \:
\mathcal{I}_t^{\ell_1 r}(\xi_{i_1},k_1)\cdots \mathcal{I}_t^{\ell_m r_m}(\xi_{i_m},k_m)
\end{align*}
so applying expectation it follows that

\begin{align*}
\esp{ \left( \sum_{\ell=1}^m a_{\ell} X_\ell \right)^n} &= \sum_{\bm{\ell}\in\mathcal{S}_{m,n}} \binom{n}{\ell_1\:\cdots\:\ell_m} a_{1}^{\ell_1}\,\cdots\,a_{m}^{\ell_m} \: M_t^{(\bm{\ell r})}(\pmb{\xi}_i;\pmb{k}).
\end{align*}

\end{proof}

\subsection{Proof of Proposition 2}\label{proof3}
With the conditions in the Proposition, it follows from Theorem 27.7 in \cite{sato} that $\pmb{\xi}(t)$ has an absolutely continuous (a.c.) distribution. We observe that for any $i\in\left\{i_1,\ldots,i_m\right\}$ then $\pmb{\xi}(t)$ is a.c. and following \cite{epilijoiprunst}, for $k\in\mathbb{N}\setminus\{0\}$,

$$
\mu_{i,t}^{(k)}=k\int_0^t e^{-\xi_i(u) }u^{k-1} \d u
$$

\noindent can be approximated by a.s. converging inferior Riemann sums which are a.c. and monotonic so $\mu_{i,t}^{(k)}$ is a.c. as well. The claim in the proposition follows as continuous transformation of a.c. r.v.'s are a.c. again.

\renewcommand{\thesection}{B}
\renewcommand{\thesubsection}{B.\arabic{subsection}}
\renewcommand{\theequation}{\Alph{subsection}.\arabic{equation}} 
\setcounter{equation}{0}
\renewcommand{\thedefinition}{B.\arabic{definition}} 
\setcounter{definition}{0}
\renewcommand{\thetheorem}{B.\arabic{theorem}} 
\setcounter{theorem}{0}
\renewcommand{\theproposition}{B.\arabic{proposition}} 
\setcounter{proposition}{0}

\section{Review of NTR multiple-sample priors and their inference procedures}\label{AppB}

\renewcommand{\thesection}{C}
\renewcommand{\thesubsection}{C.\arabic{subsection}}
\renewcommand{\theequation}{\Alph{subsection}.\arabic{equation}} 
\setcounter{equation}{0}

\begin{definition}
We say that $Y$ has a \textit{neutral to the right} (NTR) distribution given a subordinator $\xi$ if 
$$
\probc{ Y > y  }{\xi} = e^{-\xi(y)}.
$$

Furthermore for $d,n_1,\ldots,n_d\in\mathbb{N}$ we say that $\pmb{Y}_i=\left\{ Y_{i,j} \right\}_{j=1}^{n_i}$, $1\leq i\leq d$, are NTR multiple-samples given a vector of subordinators $\pmb{\xi}$ if for $j_i \in \left\{1,\ldots,n_i\right\}$, $1\leq i\leq d$, we have that

$$
\probc{ Y_{1,j_1} > y_1,\cdots,  Y_{d,j_d} > y_1 }{ \pmb{\xi} } = e^{-\xi_1(y_1)-\ldots - \xi_d(y_d)}.
$$
\end{definition}

Given $n=n_1+\ldots+n_d$ possibly censored to the right observations with respective order statistics without repetitions, $T_1<\ldots<T_k$, $k\leq n$, we denote 

\begin{align*}
n_{i,j}^{(e)}&=\#\left\{ \text{exact observations in sample }j\text{ at time } T_i \right\},
\\
n_{i,j}^{(c)}&=\#\left\{ \text{censored observations in sample }j\text{ at time } T_i \right\}
\end{align*}
and
\begin{align*}
\bar{n}_{i,j}^{(e)} = \sum_{l=j}^k  n^{(e)}_{l,j}, \quad
\bar{n}_{i,j}^{(c)} = \sum_{l=j}^k  n^{(c)}_{l,j}.
\end{align*}
to define vectors $\bar{\pmb{n}}_i^{(e)}=(\bar{n}_{i,1}^{(e)},\ldots, \bar{n}_{i,d}^{(e)})$, $\bar{\pmb{n}}_i^{(c)}=(\bar{n}_{i,1}^{(c)},\ldots, \bar{n}_{i,d}^{(c)})$;$1\leq i\leq k$, $1\leq j\leq d$. We fix $T_{0}=0$ and  $T_{k+1}=\infty$, $n_{k+1,j}^{(e)}=n_{k+1,j}^{(c)}=0$, $1\leq j\leq d$.
\\
\begin{theorem}\label{posteo}(Posterior characterization, \cite{riva2018a}.)
Let $\pmb{\mathcal{D}}$ be possibly censored to the right survival data corresponding to NTR multiple-sample with associated L\'evy measure of the form $\nu_t(\d \pmb{x})=\int_0^t \nu( \d  \pmb{x},s ) \d s$, then the posterior distribution of $\pmb{\xi}$ in $[0,T_f]$ is given by

$$
\pmb{\xi}^\circ + \sum_{ \left\{ i \, :\, T_i \leq T_f \text{ is an exact observation} \right\} } \pmb{W}_i \delta_{T_i}
$$

where $\pmb{\xi}^\circ$ is a vector of subordinators with intensity

$$
\nu^\circ_t (\mathrm{d} \pmb{x} ) = \sum_{i=1}^k \pmb{1}\left\{ t\in [T_{i-1},T_i) \right\}\int_0^t e^{- \langle \bar{\pmb{n}}^{(c)}_i+\bar{\pmb{n}}^{(e)}_i \,,\, \pmb{x} \rangle  } 
\nu( \pmb{x}, s) \d \pmb{x} \d s,
$$

$W_j\sim\mathcal{L}(f_j)$ with

\begin{align*}
f_i(\pmb{x}) = \frac{ \prod_{j=1}^d \left( e^{-\left( \bar{n}^{(c)}_{i,j} + \bar{n}^{(e)}_{i+1,j}\right)x_j }\left(1-e^{-x_i}\right)^{n_{i,j}^{(e)}} \right) \nu(\pmb{x},T_i )  }{ \int_{(\mathbb{R}^+)^d} \prod_{j=1}^d \left( e^{-\left( \bar{n}^{(c)}_{i,j}+ \bar{n}^{(e)}_{i+1,j} \right)u_j }\left(1-e^{-u_i}\right)^{n_i^{(e)}} \right) \nu(\pmb{u},T_i ) \d\pmb{u}  }
\end{align*}

and $\left\{ W_i \right\}_{i=1}^k \perp \pmb{\xi}^\circ$.
\end{theorem}
\medskip
Multiple-sample NTR priors allow for calculation of the marginal likelihood where the vector of subordinators has been integrated out.
\medskip
\begin{theorem}\label{likeliteo}(Marginal likelihood, \cite{riva2018a}.)
The likelihood for a multiple-sample NTR model as in theorem \ref{posteo} with $n$ observations and Laplace exponent denoted $\psi$ is given by
\begin{align*}
L(\mathcal{\pmb{D}}) &= e^{-\sum_{i=1}^k \left( \psi_{T_{(i)}}(\bar{\pmb{n}}^{(e)}_i +\bar{\pmb{n}}_i^{(c)}) - \psi_{T_{(i-1)}}(\bar{\pmb{n}}^{(e)}_i +\bar{\pmb{n}}_i^{(c)}) \right)}
\\ & \hphantom{=} \times \prod_{ \left\{ i \, :\, T_i  \text{ is an exact observation} \right\} } \int_{\mathbb{R}_+ } \prod_{j=1}^d \left( e^{-\left( \bar{n}^{(c)}_{i,j} + \bar{n}^{(e)}_{i+1,j}\right)x_j }\left(1-e^{-x_i}\right)^{n_{i,j}^{(e)}} \right) \nu(\pmb{x},T_i ) \mathrm{d}\pmb{x}
\end{align*}
\end{theorem}

\medskip
We observe that in the above likelihood there is parameter dependence in $\nu_t$ and the Laplace exponent.
\\
Inference for multiple-sample NTR priors is usually eased when the Laplace exponent can be computed explicitly; this can be seen by expressing the integrals in theorems \ref{posteo} and \ref{likeliteo} as difference of Laplace exponents, see for example Lemma 1 in \cite{riva2022}. The Laplace exponent of the compound subordinators discussed and used in the main manuscript can be computed using the next proposition.

\begin{proposition}(Laplace exponent for compound vector of subordinators, \cite{riva2021}.) \label{propB}
The multivariate Laplace exponent $\psi$ for a compound vector of subordinators with score distributions $H$ and directing L\'evy measure $\nu^\star$ with associated univariate Laplace exponent $\psi^\star$ is given by

$$
\psi_t(\pmb{\lambda})=\mathbb{E}_{\pmb{Z}\sim H}\left[ \psi^\star\left( \lambda_1Z_i+\ldots +\lambda_dZ_d\right)\right].
$$
\end{proposition}

\section{Beta-Stacy process moments}\label{AppC}

In this appendix section we present the necessary formulas for implementation of the moment calculations needed for performing density estimation as presented in Algorithm 1 of the main work when using the $(\pmb{\pi}_1,\pmb{\pi}_2)$--stratified-compound Log-Beta process introduced in section 3 of the main work. We begin by focusing on the univariate Log-Beta process corresponding to the Beta-Stay prior and the corresponding formulas for implementing the recursion of Theorem 1 in the main work. Then we present the appropriate formulas for the two multiple samples setting. For simplicity's sake, in the following calculations we have assumed that there are no repeated observations values in the survival data; the alternative case is similarly handled performing binomial theorem expansions.
\\ \\ \noindent Let $\gamma$ be a positive integer, $\alpha(s)$ a probability density function in $\re_+$ and $\beta(s)=\int_s^\infty \alpha(s)\d s$. The associated Log-Beta process has L\'evy intensity given by

\begin{align*}
\nu_t(\d x)= \int_0^t \nu(\d x, \d s)= \frac{1}{1-e^{-x}}\int_0^t e^{-\gamma \beta(s) x}\gamma\alpha(s )\d s\d x,
\end{align*}

see \cite{betastacy}.

The corresponding a priori Laplace exponent evaluated at an integer $m\geq 1$ is given by

\begin{align*}
\psi_t(m) & =
\int_0^\infty (1-e^{- m x})\nu_t(\d x)=
\int_0^\infty  \frac{1-e^{-m x}}{1-e^{- x}}
\int_0^t e^{-\gamma \beta(s)x }\gamma  \alpha(s) \d s \d x
\\ &  = \int_0^\infty  \left( \sum_{i=0}^{m-1} e^{-ix} \right)
\eval{ \left( \frac{ e^{-\gamma \beta(s) x } } {x} \right) }{0}{t} \d x
= \sum_{i=0}^{m-1} \int_0^\infty   
 \frac{ e^{-\left(\gamma\beta(t) +i\right)x } - e^{-\left(\gamma+i\right)x } }{x}  \d x
\\ & = \sum_{i=0}^{m-1}  -\log\left( \frac{ \gamma\beta(t) + i }{\gamma+i} \right)= 
\sum_{i=0}^{m-1}  \log\left( \frac{\gamma+i}{ \gamma\beta(t) + i } \right)
\end{align*}
where we recognized a Frullani integral.

If we consider observations without repetitions $T_1,\ldots , T_n$ and $T_1^{(e)},\ldots , T_{n_e}^{(e)}$ the $n_e$ exact observations, which without loss of generality we take to be sorted, then a posteriori we have a Beta-Stacy process with intensity

\begin{align*}
\nu_t^\circ(\d x)=\int_0^t e^{-R(s)x}\nu(\d x,\d s)= \frac{1}{1-e^{-x}}\int_0^t e^{-(\gamma \beta(s)+R(s)) x}\gamma \alpha(\d s)
\end{align*}

for $R(s)=\sum_{i=1}^n \mathds{1}_{[s,\infty)}(T_i)$ and fixed location jumps at $\left\{T_i^{(e)}\right\}_{i=1}^{n_e}$ with random weights $J_i\sim \mathcal{L}(f_i)$ determined by probability density function

\begin{align*}
f_i(x) \propto e^{-(R(T_i^{(e)})-1)x}(1-e^{-x})\restr{ \frac{\partial}{\partial t}\nu_t(x)}{t=T_i^{(e)}}
& = e^{-(R(T_i^{(e)})-1)x} e^{-\gamma \beta( T_i^{(e)} )x } \gamma \alpha( T_i^{(e)}  )
\\ & \propto e^{-(\gamma \beta(T_i^{(e)}) + R(T_i^{(e)})-1)x}
\end{align*}

So $J_i \sim \text{Exp}\left( \gamma \beta(T_i^{(e)})+R\left( T_i^{(e)} \right)-1 \right)$. The corresponding Laplace exponent a posteriori is $$\psi_{t|D}(u)=\psi_t^\circ - \sum_{i\, : \, T_i^{(e)}\leq t}\log\left( \esp{e^{-uJ_i}} \right)$$
with 
\begin{align*}
\psi_t^\circ(u) & =
\int_0^\infty (1-e^{- u x}) \nu_t^\circ(\d x)=
\int_0^\infty \left( \frac{1-e^{-u x}}{1-e^{- x}} \right)
\sum_{i=0}^{n_t} 
\int_{T_i}^{T_{i+1}} e^{-(\gamma \beta(s) + R(T_i) ) x }\gamma \alpha(s) \d s \d x
\\ & = \sum_{i=0}^{n_t} \int_0^\infty 
\left( \frac{1-e^{-u x}}{1-e^{- x}} \right) 
\frac{ e^{-(\gamma \beta( T_{i+1} ) + R(T_i) )x } - e^{- ( \gamma \beta(T_i) + R(T_i) )x}  }{x}  \d x
\end{align*}
where $T_{n_t}$ is the last observation which happened before time time $t$ and we have set $T_{0}=0$, $T_{n_t+1}=t$. On the other hand

\begin{align*}
-\sum_{i\, : \, T_i^{(e)}\leq t}\log\left( \esp{e^{-uJ_i}} \right)=
\sum_{i\, : \, T_i^{(e)}\leq t}\log\left( \frac{ \gamma \beta( T_i^{(e)} ) + R(T_i^{(e)})-1 + u}{ \gamma \beta( T_i^{(e)} ) + R(T_i^{(e)})-1 } \right)
\end{align*}

If we evaluate in an integer $m\geq 1$ then we have that

\begin{align*}
\psi_t^\circ(m) & = \sum_{i=0}^{n_t} \int_0^\infty  \left( \sum_{j=0}^{m-1} e^{-jx} \right)
\eval{ \left( \frac{ e^{-( \gamma\beta(s) + R(T_i) )x } }{x} \right) }{T_i}{T_{i+1}} \d x
\\ & = \sum_{i=0}^{n_t} \sum_{j=0}^{m-1}  \int_0^\infty  
 \frac{ e^{-(\gamma\beta( T_{i+1} ) + n - i + j)x } - e^{-(\gamma \beta( T_i ) + n - i +j )x}  }{x}  \d x
\\ & =  \sum_{i=0}^{n_t} \sum_{j=0}^{m-1} - \log\left( \frac{ \gamma \beta( T_{i+1} ) + n-i+j }{ \gamma\beta( T_i ) + n-i+j }   \right)=  \sum_{i=0}^{n_t} \sum_{j=0}^{m-1}  \log\left( \frac{ \gamma\beta( T_i ) + n-i+j }{ \gamma\beta( T_{i+1} ) + n-i +j}  \right)
\\ & = \left( \sum_{i=1}^{n_t}   \log\left( \frac{ \gamma\beta( T_i ) + n-i}{ \gamma\beta( T_i ) + n-i +m}  \right) \right)+ \left( \sum_{j=0}^{m-1} \log\left( \frac{ \gamma\beta( T_0 ) + n+j }{ \gamma \beta( T_{n_t+1} ) + n-n_t +j}\right)  \right),
\end{align*}

The corresponding mean survival estimate is given by considering $m=1$ above

\begin{align*}
& \hat{S}(t) = \espc{e^{-\xi_t}}{D}=e^{-\psi_t^\circ (1)}
=\prod_{i=0}^{n_t}\left(\frac{ \gamma\beta( T_{i+1} ) + n-i }{ \gamma\beta( T_i ) + n-i} \right)\prod_{j\, : \, T_j^{(e)}\leq t} \left(
\frac{ \gamma\beta( T_j^{(e)} ) + R(T_j^{(e)})- 1 }{ \gamma\beta( T_j^{(e)} ) + R(T_j^{(e)}) }
\right)
\\ & =
\frac{\gamma\beta(t)+n-n_t}{\gamma + n} 
\prod_{i=1}^{n_t}\left(\frac{ \gamma\beta( T_i ) + n-i +1 }{ \gamma\beta( T_i ) + n-i} \right)
\prod_{j\, : \, T_j^{(e)}\leq t} \left(
\frac{ \gamma\beta( T_j^{(e)} ) + R(T_j^{(e)})- 1 }{ \gamma\beta( T_j^{(e)} ) + R(T_j^{(e)}) }
\right).
\end{align*}

Observe that if all observations are exact then

\begin{align*}
\hat{S}(t) & = \frac{\gamma\beta(t)+n-n_t}{\gamma + n}  \prod_{i=1}^{n_t} \frac{ \left(
\gamma\beta( T_i ) + n-i + 1 \right)\left( 
\gamma\beta( T_i ) + n-i 
\right) }{ \left( 
\gamma\beta( T_i ) + n-i \right)\left( 
\gamma\beta( T_i ) + n-i + 1
\right)}=\frac{\gamma\beta(t)+n-n_t}{\gamma + n} ,
\end{align*}

which corresponds to the Dirichlet process case.

With the convention that $R(T_{n+1})=0$, the likelihood is given by

\begin{align*}
L(D) & = e^{-\sum_{i=0}^{n-1}  \int_0^\infty \int_{T_i}^{T_{i+1}} \left( 1-e^{- R(T_{i+1}) x} \right)  \nu(\d x, \d s ) }
\prod_{i=1}^{n_e} \int_0^\infty 
e^{- (R(T_{i}^{(e)})-1)  x }(1-e^{-x}) \restr{ \frac{\partial}{\partial t}\nu_t(x_1,x_2) }{t=T_i^{(e)}} 
\\
& = \left( \prod_{i=0}^{n-1} \prod_{j=0}^{R(T_{i+1})-1} \frac{\gamma\beta(T_{i+1})+j}{\gamma\beta(T_i)+j}\right)
\left( \prod_{i=1}^{n_e}  \gamma \alpha(T_i^e) \int_0^\infty e^{- (\gamma \beta(T_i^e)+R(T_{i}^{(e)})-1)  x }\d x \right)
\\
& = \left( \prod_{j=0}^{n-1} \frac{\gamma\beta(T_{i+1})+R(T_{i+1})-1}{\gamma +j}\right)
\left( \prod_{i=1}^{n_e} \frac{ \gamma \alpha(T_i^e) }{ \gamma \beta(T_i^e)+R(T_{i}^{(e)})-1 } \right)
\end{align*}

We observe that if all observations are exact then 

\begin{align*}
L(D) & = \left( \prod_{j=0}^{n-1} \frac{\gamma\beta(T_{i+1})+R(T_{i+1})-1}{\gamma +j} \right)
\left( \prod_{i=0}^{n-1} \frac{ \gamma \alpha(T_{i+1}) }{ \gamma \beta(T_{i+1})+R(T_{i+1})-1 } \right)
\\ &=  \prod_{j=0}^{n-1} \frac{\gamma\alpha(T_{i+1}) }{\gamma +j}
\end{align*}
which tends to the likelihood associated to the base density $\alpha$ as the precision parameter  $\gamma\to\infty$.
\\
\\
We denote $\xi\,|\,\mathcal{D}$ as the posterior subordinator given the survival data. To use the recursion of Theorem 1 in the one sample case we need basically to evaluate


\begin{align*}
e^{ -\left( \psi_{t}^\circ (m) - \psi_{t\,|\,D}(m-1) \right) }
& = \frac{\gamma\beta(t)+n-n_t+m-1}{\gamma+n+m-1} \prod_{i=1}^{n_t}
\frac{ \gamma\beta(T_i)+n-i + m }{ \gamma\beta(T_i)+n-i + m-1 }
\\ & \hphantom{=} \times
\prod_{i\, : \, T_i^{(e)}\leq t} \left(
\frac{ \gamma\beta( T_i^{(e)} ) + R( T_i^{(e)}) + m - 2 }{ \gamma\beta( T_i^{(e)} ) + R( T_i^{(e)}) + m-1 } \right) 
\end{align*}

so

\begin{align*}
& M_{s,t}^{(m)}(\xi \, |\, D;k)
 = m k \int_s^t  M_{v,t}^{(m-1)}(\xi \, |\, D;k)e^{ -\left( \psi_{v\, |\, D}(m) - \psi_{v\, |\, D}(m-1) \right) } v^{k-1} \d v 
\\ & = m k \int_s^t  M_{v,t}^{(m-1)}(\xi \, |\, D;k) 
\prod_{i=0}^{n_v}\left(\frac{e^{-\lambda T_{i+1} } + n-i + m-1 }{e^{-\lambda T_i } + n-i + m-1} \right)  \\&\quad\quad\times \prod_{i\, : \, T_i^{(e)}\leq v} \left(
\frac{ \lambda e^{-\lambda T_i^{(e)}} + R(T_i^{(e)})- 1 + m - 1 }{ \lambda e^{-\lambda T_i^{(e)}} + R( T_i^{(e)}) - 1 + m } \right)v^{k-1}
\d v .
\end{align*}

Before presenting the analogue quantities for implementation of the recursion in Theorem 1 for the two samples case, we show formulas for a univariate compound random measure which as discussed in section two of the main work corresponds to multiplying the weights of a subordinator by i.i.d. variables with law corresponding to a v.a., lets say $Z$. In particular we will consider $Z$ a non negative integer valued random variable with probability mass function $h$. A univariate compound subordinator with $\mathcal{L}(h)$ score distribution and Beta-Stacy directing L\'evy measure has a priori Laplace exponent, see Proposition \ref{propB}, which evaluated at integer $m\geq 1$ values is given by

\begin{align*}
\psi^\text{Z-Comp}_t(m)
=\esp{\psi_t(mZ)}
= \esp{ \psi_t(mZ) }
= \esp{ \sum_{i=0}^{mZ-1}  \log\left( \frac{\gamma+i}{ \gamma \beta(t) + i } \right) }´
\end{align*}

and a posteriori is given by

\begin{align*}
& \psi_t^{\text{Z-Comp}\, \circ}(m) =
\sum_{i=0}^{n_t} \int_0^\infty \int_{T_i}^{T_{i+1}} (1-e^{-m x})e^{-(n-i)x}\nu^{\text{Z-Comp}}(\d x, \d s )
\\ &= 
\sum_{i=0}^{n_t} \int_0^\infty \int_{T_i}^{T_{i+1}} \sum_{z=0}^\infty (1-e^{-m z x})e^{-(n-i)z x}h(z) \nu( \d x, \d s )
\\
 &= 
\esp{ \sum_{i=0}^{n_t} \int_0^\infty \int_{T_i}^{ T_{i+1}}  \left( \frac{1-e^{-m Z x} }{1-e^{-x}}\right)e^{-\left( \gamma\beta(s) + (n-i)Z \right)x } \gamma \alpha(s) \d s \d x }
\\
&= 
\esp{ \mathds{1}_{\left\{ Z>0 \right\} }  \sum_{i=0}^{n_t}  \sum_{j=0}^{mZ-1}
\int_0^\infty \frac{ e^{-\left(\gamma\beta( T_{i+1}) + (n-i)Z + j \right) x } - e^{-\left( \gamma\beta( T_i) + (n-i)Z + j \right)x } }{x} \d x }
\\
& = \mathbb{ E }\left[ \mathds{1}_{\left\{ Z>0 \right\} } 
\sum_{i=0}^{n_t}  \sum_{j=0}^{mZ-1}
\log\left( \frac{ 
\gamma\beta( T_i) + (n-i)Z + j }{
\gamma\beta( T_{i+1}) + (n-i)Z + j  }
\right) \right]
\end{align*}

%

The fixed location jumps at $\left\{T_i^{(e)}\right\}_{i=1}^{n_e}$ have random weights $J_i$ with density determined by

\begin{align*}
f_i(x) & \propto e^{-(R(T_i^{(e)})-1)x}(1-e^{-x})\restr{ \frac{\partial}{\partial t}\nu_t^{\text{Z-Comp}}(x)}{t=T_i^{(e)}}
\end{align*}

so 

\begin{align*}
\mathbb{E}\left[ e^{- u J_i}\right] &=  \frac{ \mathbb{E}\left[ \int_0^\infty e^{- \left( u + R( T_i^{(e)}) - 1\right) xZ}(1-e^{-xZ}) \restr{ \frac{\partial}{\partial t}\nu_t(x) }{t= T_i^{(e)}} \mathrm{d}x \right] }{ \mathbb{E}\left[ \int_0^\infty e^{-(R( T_i^{(e)})-1)xZ}(1-e^{-xZ}) \restr{ \frac{\partial}{\partial t}\nu_t(x) }{t= T_i^{(e)}} \mathrm{d}x \right] }
\\&=  \frac{ \mathbb{E}\left[ \int_0^\infty e^{- \left( u + R(T_i^{(e)}) - 1\right)xZ}\left( \frac{1-e^{-xZ}}{1-e^{-x}}\right) e^{- \gamma\beta(T_i^{(e)})x } \d x  \right] }{ \mathbb{E}\left[ \int_0^\infty e^{-(R( T_i^{(e)})-1)xZ} \left( \frac{1-e^{-xZ}}{1-e^{-x}}\right)e^{-\gamma\beta( T_i^{(e)})x } \d x \right] }
\\&=  \frac{ \mathbb{E}\left[ \sum_{j=0}^{Z-1} \int_0^\infty e^{-\left( u + R( T_i^{(e)})-1)\right)
xZ}e^{-jx}e^{-\gamma \beta( T_i^{(e)} ) x} \d x \right] }{ \mathbb{E}\left[ \sum_{j=0}^{Z-1} \int_0^\infty e^{-(R( T_i^{(e)})-1)xZ} e^{-jx} e^{-\gamma \beta( T_i^{(e)})x } \d x \right] }
\\  & = \frac{ \mathbb{E}\left[ \sum_{j=0}^{Z-1} \frac{1}{ (u+R( T_i^{(e)})-1)Z+j+\gamma\beta( T_i^{(e)}) } \right] }{ \mathbb{E}\left[ \sum_{j=0}^{Z-1} \frac{1}{(R( T_i^{(e)})-1)Z+j+\gamma \beta( T_i^{(e)}) } \right] }
\end{align*}
%
%
The posterior Laplace exponent can be evaluated analytically via the two previous expectation calculation which correspond to a discrete r.v..
\\
\\
\noindent If we consider two-sample observations $T_{1,1},\ldots T_{1,n_1},T_{2,1},\ldots , T_{2,n_2}$ without repetitions, the sorting of the union of the observations across samples $T_1,\ldots , T_n$ with $n=n_1+n_2$ and the sorting of the union of exact observations across samples $T_1^{(e)},\ldots , T_{n_e}^{(e)}$. Let $R_j(s)=\sum_{i=1}^{n_j} \mathds{1}_{[s,\infty)}(T_{j,i})$, $j\in\left\{ 1,2\right\}$ and $\pmb{Z}=(Z_1,Z_2)$ be a random vector with positive integer entries. A bivariate compound vector of subordinators with score distribution $\mathcal{L}(\pmb{Z})$, Beta-Stacy directing L\'evy measure a priori has Laplace exponent such that for integers $m_1,m_2\geq 1$

\begin{align*}
\psi_t^{\pmb{Z}\text{-Comp}}(m_1,m_2)&=\esp{\psi_t(m_1Z_1 + m_2 Z_2)}
 = \esp{ \sum_{i=0}^{m_1 Z_1 +m_2 Z_2-1}  \log\left( \frac{1+i}{ e^{-\lambda t} + i } \right) }
\end{align*}
and a posteriori
\begin{align*}
& \psi_t^{\pmb{Z}\text{-Comp}\, \circ}(m_1,m_2) = 
\int_0^\infty \int_0^\infty \left( 1-e^{-m_1 x_1 -m_2 x_2} \right)\nu_t^{\pmb{Z}\text{-Comp}\, \circ} (x_1, x_2) \d x_1  \d x_2
\\ &= 
\sum_{i=0}^{n_t} \int_0^\infty \int_0^\infty \int_{T_i}^{T_{i+1}} \left( 1-e^{-m_1 x_1 - m_2x_2} \right)e^{ -R_1 (T_{i+1})x_1 - R_2 (T_{i+1})x_2 } \nu^{\pmb{Z}\text{-Comp}}(\d x_1, \d x_2, \d s )
\\ &= 
\sum_{i=0}^{n_t} \int_0^\infty \int_{ T_i}^{ T_{i+1}} \sum_{z=0}^\infty(1-e^{-(m_1 z_1  x + m_2 z_2) x})e^{ -( R_1 ( T_{i+1}) z_1 - R_2 (T_{i+1})z_2)x } h(z_1,z_2) \nu( \d x, \d s )
\\
 &= 
\esp{ \sum_{i=0}^{n_t} \int_0^\infty \int_{ T_i}^{ T_{i+1}}  \left( \frac{1-e^{-( m_1 Z_1 +m_2 Z_2) x} }{1-e^{-x}}\right)e^{-\left( \gamma\beta(s) + R_1 ( T_{i+1})Z_1 + R_2( T_{i+1})Z_2 \right)x } \gamma \alpha(s) \d s \d x }
\\
&= 
\mathbb{E}\left[ \mathds{1}_{\left\{ Z_1+Z_2>0 \right\}} \sum_{i=0}^{n_t}  \sum_{j=0}^{m_1Z_1+m_2Z_2-1}
\int_0^\infty \left( \frac{ e^{-\left(\gamma\beta( T_{i+1})  + R_1 ( T_{i+1})Z_1 + R_2 ( T_{i+1})Z_2  + j \right) x }}{x} \right. \right. \\ & \left. \left. \quad - \frac{e^{-\left( \gamma\beta( T_i) + R_1 ( T_{i+1})Z_1 + R_2 (T_{i+1})Z_2  + j \right)x } }{x} \d x\right) \right]
\\
& = \mathbb{ E }\left[  \mathds{1}_{\left\{ Z_1+Z_2>0 \right\}}  \sum_{i=0}^{n_t}  \sum_{j=0}^{m_1Z_1+m_2Z_2-1}
\log\left( \frac{ 
\gamma\beta( T_i) + R_1 ( T_{i+1})Z_1 + R_2 ( T_{i+1})Z_2  + j }{
\gamma\beta( T_{i+1}) + R_1 ( T_{i+1})Z_1 + R_2 ( T_{i+1})Z_2  + j  }
\right) \right]
\end{align*}

where sum over an empty set is taken to be zero.
%
Analogously, if $m_k=1$ and $m_{-k+3}=0$, $k\in\left\{ 1,2\right\}$, then the marginal Laplace exponents a posteriori are given by

\begin{align*}
& \psi_t^{\pmb{Z}\text{-Comp}\, \circ(k)}(m_k) = 
\esp{ \mathds{1}_{\left\{ Z_k >0 \right\} }  \sum_{i=0}^{n_t}  \sum_{j=0}^{m_kZ_k-1}
\log\left( \frac{ 
\gamma\beta(T_i) + R_1(T_{i+1})Z_1 +R_2(T_{i+1})Z_2  + j }{
\gamma\beta(T_{i+1}) + R_1(T_{i+1})Z_1 +R_2(T_{i+1})Z_2  + j  }
\right) }
\end{align*}

Let $\delta_i^{(j)}$ be the indicator variable corresponding to observation $i$ being exactly observed in population $1\leq i \leq n$, $1\leq j \leq 2$. The fixed location jumps at $\left\{ T_i^{e}\right\}_{i=1}^{n_e}$ with random weights $\pmb{J}_i=(J_{1,i},J_{2,i})\sim \mathcal{L}(f_i)$ are determined by

\begin{align*}
f_i(x_1,x_2) & \propto e^{-(R_1( T_i^{(e)})-1)x_1-(R_2( T_i^{(e)})-1)x_2} 
(1-e^{-x_1})^{\delta_i^{(1)}} (1-e^{-x_2})^{\delta_i^{(2)}} \frac{\partial}{\partial t}\nu_t^{\pmb{Z}\text{-Comp}}(x_1,x_2).
\end{align*}.

The corresponding Laplace transform for the random weights $\pmb{J}_i$  evaluated at integers $m_1,m_2$ becomes

\begin{align*}
& \esp{e^{-m_1J_1-m_2J_2}}  = \\ & \frac{ \int_0^\infty \int_0^\infty 
e^{-\left( R_1(T_i^{(e)}) -\delta_i^{(e,1)} + m_1\right)x_1- \left( R_2(T_i^{(e)})-\delta_i^{(e,2)}+ m_2\right) x_2}(1-e^{-x_1})^{\delta_i^{(e,1)}} (1-e^{-x_2})^{\delta_i^{(e,2)}} \restr{ \frac{\partial}{\partial t}\nu_t^{\pmb{Z}\text{-Comp}}(x_1,x_2) }{t=T_i^{(e)}} }{
 \int_0^\infty \int_0^\infty 
e^{- \left( R_1(T_i^{(e)}) -\delta_i^{(e,1)} \right) x_1- \left( R_2(T_i^{(e)}) - \delta_i^{(e,2)} \right) x_2}(1-e^{-x_1})^{\delta_i^{(1)}} (1-e^{-x_2})^{\delta_i^{(2)}} \restr{ \frac{\partial}{\partial t}\nu_t^{\pmb{Z}\text{-Comp}}(x_1,x_2) }{t=T_i^{(e)}} 
} \\ & = \frac{ \esp{ \int_0^\infty 
e^{- \left( ( R_1(T_i^{(e)}) -\delta_i^{(e,1)} + m_1 )Z_1 +  ( R_2(T_i^{(e)}) -\delta_i^{(e,2)} + m_2)Z_2 \right) x } \frac{ (1-e^{-Z_1x})^{\delta_i^{(e,1)}} (1-e^{-Z_2x})^{\delta_i^{(e,2)}} }{(1-e^{-x})} e^{-\gamma\beta(T_i^{(e)})x } \d x } 
}{ \esp{  \int_0^\infty  e^{-\left( ( R_1(T_i^{(e)})  -\delta_i^{(e,1)})Z_1 + (R_2(T_i^{(e)}) -\delta_i^{(e,2)})Z_2\right)x } \frac{ (1-e^{-Z_1x})^{\delta_i^{(e,1)}} (1-e^{-Z_2x})^{\delta_i^{(e,2)}} }{(1-e^{-x})} e^{-\gamma\beta(T_i^{(e)})x } \d x } }
\\ & = \frac{ \esp{ \sum_{j=0}^{ Z_1\delta_i^{(e,1)} + Z_2\delta_i^{(e,2)} -1 } \frac{1}{  ( R_1(T_i^{(e)}) -\delta_i^{(e,1)} + m_1 )Z_1 +  ( R_2(T_i^{(e)}) -\delta_i^{(e,2)} + m_2)Z_2 + j + \gamma\beta(T_i^{(e)})} \; } 
}{ \esp{ \sum_{j=0}^{ Z_1\delta_i^{(e,1)} + Z_2\delta_i^{(e,2)} -1 } \frac{1}{  ( R_1( T_i^{(e)}) -\delta_i^{(e,1)}) Z_1 + ( R_2( T_i^{(e)}) -\delta_i^{(e,2)}) Z_2 + j + \gamma\beta( T_i^{(e)})} \; }  }
\end{align*}

\noindent So the Laplace exponent a posteriori at integer values is given by

$$
\psi_{t\,|\, D}^{\pmb{Z}\text{-Comp}}(m_1,m_2)= \psi_t^{\pmb{Z}\text{-Comp}\, \circ}(m_1,m_2) - \sum_{i\, : \, T_i^{(e)}\leq t}\log\left( \esp{e^{-m_1 J_i-m_2 J_2}} \right)
$$
is readily available.
\\
%
%

The likelihood is given by

\begin{align*}
& L(\pmb{v};D)  = e^{-\sum_{i=0}^{n-1} \int_0^\infty \int_0^\infty \int_{T_i}^{T_{i+1}} \left( 1-e^{- R_1 (T_{i+1}) x_1 - R_2 (T_{i+1}) x_2} \right)  \nu^{\pmb{Z}\text{-Comp}}(\d x_1, \d x_2, \d s ) }
\\ & \times 
\prod_{i=1}^{n_e} \int_0^\infty \int_0^\infty 
e^{- \left( R_1(T_i^{(e)}) -\delta_i^{(e,1)} \right) x_1- \left( R_2(T_i^{(e)}) - \delta_i^{(e,2)} \right) x_2}(1-e^{-x_1})^{\delta_i^{(1)}} (1-e^{-x_2})^{\delta_i^{(2)}} \restr{ \frac{\partial}{\partial t}\nu_t^{\pmb{Z}\text{-Comp}}(x_1,x_2) }{t=T_i^{(e)}} 
\\
& =
\exp\left( -\mathbb{E}_{\pmb{v}}\left[ \mathds{1}_{\left\{ Z_1+Z_2 >0 \right\} }  \sum_{i=0}^{n-1}  \sum_{j=0}^{ R_1(T_{i+1})Z_1+R_2(T_{i+1})Z_2-1 }
\log\left( \frac{ 
\gamma\beta(T_i) +  j }{
\gamma\beta(T_{i+1}) + j  }
\right) \right] \right)
\\ & \times 
\prod_{i=1}^{n_e} \mathbb{E}_{\pmb{v}} \left[ \sum_{j=0}^{ Z_1\delta_i^{(1)} + Z_2\delta_i^{(2)} -1 } \frac{\gamma\beta(T_i^{(e)})}{ \gamma\beta(T_i^{(e)}) + \left( R_1(T_i^{(e)}) -\delta_i^{(e,1)}\right) Z_1 + \left( R_2(T_i^{(e)}) -\delta_i^{(e,2)} \right) Z_2 + j }  \right]
\end{align*}

Finally the recursion from Theorem \ref{teo1} is determined by the quantities

\begin{align*}
M_{s,t}^{(1,0)}( \pmb{\xi} \, | \, D;\bm{k}) &= k_1 \int_s^t e^{-\psi^{ \pmb{Z} \text{-Comp}\, \circ}_v (1,0)} v^{k_1-1} \d v =   k_1 \int_s^t \hat{S}_1(v) v^{k_1 - 1} v^{k_1-1} \d v,
\\
M_{s,t}^{(0,1)}( \pmb{\xi} \, | \, D;\pmb{k}) &= k_2\int_s^t e^{-\psi^{ \pmb{Z}\text{-Comp}\, \circ}_v (0,1)} v^{k_2-1} \d v =   k_2\int_s^t \hat{S}_2(v) v^{k_2-1}\d v.
\end{align*}

and for $m_1>m_3>0,\,m_2>m_4>0$

\begin{align*}
&e^{-\left( \psi_{t\, |\, D}(m_1,m_2) - \psi_{t\, | \, D}(m_3,m_4)\right) }  = 
\\ & \exp\left( 
\esp{  \mathds{1}_{\left\{  Z_1+ Z_2>0 \right\}} \sum_{i=0}^{n_t}  \sum_{j=m_3 Z_1+m_4 Z_2}^{m_1Z_1+m_2Z_2-1} \log\left( \frac{ 
\gamma\beta( T_i) + R_1 ( T_{i+1})Z_1 + R_2 ( T_{i+1})Z_2  + j }{
\gamma\beta( T_{i+1}) + R_1 ( T_{i+1})Z_1 + R_2 ( T_{i+1})Z_2  + j  }
\right)  } \right)
\\ &  \times \prod_{i\, : \, T_i^{(e)}\leq t} \left(
\frac{ \esp{ \sum_{j=0}^{ Z_1\delta_i^{(e,1)} + Z_2\delta_i^{(e,2)} -1 } \frac{1}{  ( R_1(T_i^{(e)}) -\delta_i^{(e,1)} + m_1 )Z_1 +  ( R_2(T_i^{(e)}) -\delta_i^{(e,2)} + m_2)Z_2 + j + \gamma\beta(T_i^{(e)})} \; } 
}{ \esp{ \sum_{j=0}^{ Z_1\delta_i^{(e,1)} + Z_2\delta_i^{(e,2)} -1 } \frac{1}{  ( R_1(T_i^{(e)}) -\delta_i^{(e,1)} + m_3 )Z_1 +  ( R_2(T_i^{(e)}) -\delta_i^{(e,2)} + m_4 )Z_2 + j + \gamma\beta(T_i^{(e)})} \; } }
\right)
\end{align*}

and for $m_1>m_2>0$

\begin{align*}
&e^{-\left( \psi_{t\, |\, D}(m_1,0) - \psi_{t\, | \, D}(m_2,0)\right) }  = \\ & \exp\left( 
\mathbb{ E }\left[ \mathds{1}_{\left\{ Z_k >0 \right\} }  \sum_{i=0}^{n_t}  \sum_{j=m_2 Z_2}^{m_1 Z_1-1}
\log\left( \frac{ 
\gamma\beta(T_i) + R_1(T_{i+1})Z_1 + R_2(T_{i+1})Z_2  + j }{
\gamma\beta(T_{i+1}) + R_1(T_{i+1})Z_1 + R_2(T_{i+1})Z_2  + j  }
\right)  \right. \right.
\\ &  \times \prod_{i\, : \, X_i^{(e)}\leq t} \left(
\frac{ \esp{ \sum_{j=0}^{ Z_1\delta_i^{(e,1)} + Z_2\delta_i^{(e,2)} -1 } \frac{1}{  ( R_1(X_i^{(e)}) -\delta_i^{(e,1)} + m_1 )Z_1 +  ( R_2(X_i^{(e)}) -\delta_i^{(e,2)} )Z_2 + j + \gamma\beta(X_i^{(e)})} \; } 
}{ \esp{ \sum_{j=0}^{ Z_1\delta_i^{(e,1)} + Z_2\delta_i^{(e,2)} -1 } \frac{1}{  ( R_1(X_i^{(e)}) -\delta_i^{(e,1)} + m_2 )Z_1 +  ( R_2(X_i^{(e)}) -\delta_i^{(e,2)} )Z_2 + j + \gamma\beta(X_i^{(e)})} \; } }
\right)
\end{align*}
with analogous formulas for the other marginal case where the Laplace exponent difference corresponds to evaluations at $(0,m_1)$ and $(0,m_2)$.

\section{Variance correlation}\label{AppD}

Using the moment formulas of Theorem \ref{teo1} in the main work we obtain that 

\begin{align*}
&\text{Corr}\left(\sigma^2_{1,t},\sigma^2_{2,t}\right)=   \left( M_t^{(1,1)}(\pmb{\xi};\pmb{2}) - M_t^{(1,2)}\left(\pmb{\xi};(2,1)\right) - M_t^{(2,1)}\left(\pmb{\xi};(1,2)\right) + M_t^{(2,2)}(\pmb{\xi};\pmb{1}) \right.
\\ & \left. - \left( M_t^{(1)}(\xi_1;2) - M_t^{(2)}(\xi_1;1) \right)\left( M_t^{(1)}(\xi_2;2) - M_t^{(2)}(\xi_2;1) \right) \right)\left[ \left( M^{(2)}_t(\xi_1;2) -2M_t^{(1,2)}\left((\xi_1,\xi_1);(2,1)\right)
 \right. \right.
\\ & \left. + M_t^{(4)}(\xi_1;1) -\left( M_t^{(1)}(\xi_1;2) - M_t^{(2)}(\xi_1;1) \right)^2 \right)\left(  M^{(2)}_t(\xi_2;2) -2M_t^{(1,2)}\left((\xi_2,\xi_2);(2,1)\right) + 
M_t^{(4)}(\xi_2;1) \right.
\\ & \left. - \left( M_t^{(1)}(\xi_2;2) - M_t^{(2)}(\xi_2;1) \right)^2
\right]^{-\frac{1}{2}}
\end{align*}

\begin{center}
\begin{figure}[h!]
\begin{center}
\begin{tabular}{cc}
\includegraphics[width=0.49\linewidth]{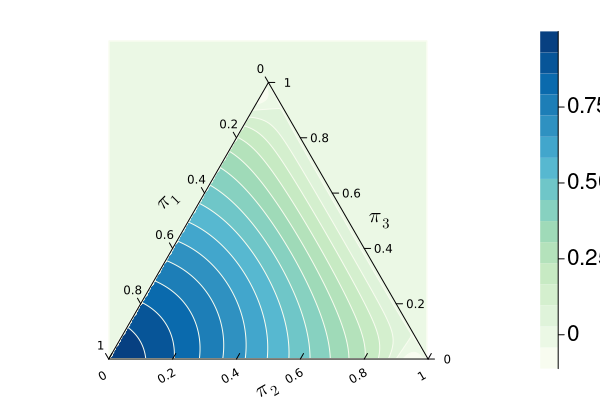} &
\includegraphics[width=0.49\linewidth]{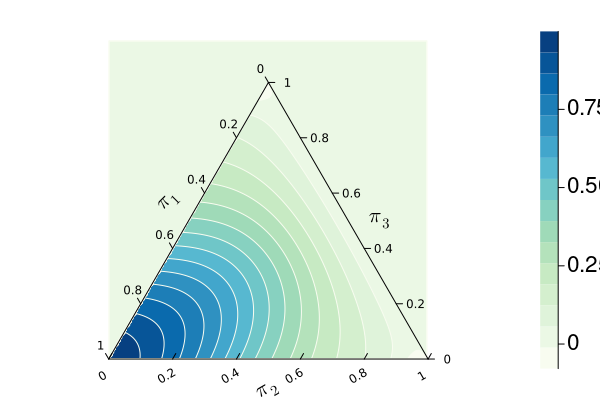} \\
(i) & (ii)
\end{tabular}
\caption{Correlation between restricted variances $\sigma_{1,t}^2,\sigma_{1,t}^2$ for (i) $t=5$ and (ii) $t=10$ in $\bm{\pi}$--compound Beta-Stacy model.} 
\label{varcorrPlot}
\end{center}
\end{figure}
\end{center}

We show in Figure \ref{varcorrPlot} such correlation for the $\pmb{\pi}$--compound Beta-Stacy process for different values of $\pmb{\pi}$.

\end{document}

%% file: definitions.tex
\newcommand{\bb}[1]{\mathbb{#1}}

\renewcommand{\d}{\ensuremath{\mathrm{d}}}

\newcommand{\sip}{\bb{P}}

\newcommand{\se}{\ensuremath{\bb{E}}}

\newcommand{\re}{\ensuremath{\mathbb{R}}}

\newcommand{\prob}[1]{\ensuremath{\sip\! \left[ #1 \right]}}

\newcommand{\probc}[2]{\ensuremath{\sip\! \left[ #1 \, | #2 \right]}}

\newcommand{\esp}[1]{\ensuremath{\se\! \left[ #1 \right]}}
\newcommand{\espc}[2]{\ensuremath{\se\! \left[ #1 | #2 \right]}}





%% file: Arxiv_Final_Submission.bbl
\begin{thebibliography}{7}
\expandafter\ifx\csname natexlab\endcsname\relax\def\natexlab#1{#1}\fi

\bibitem[Antoniak (1974)]{antoniak}
Antoniak, C.A. (1974). Mixtures of Dirichlet processes with application to Bayesian nonparametric problems. \textit{Ann. Statist.} \textbf{2}, 1152--1174.

\bibitem[Bertoin and Yor (2005)]{bertoinyor}
Bertoin, J., and Yor, M. (2005). Exponential functionals of Lévy processes. \textit{Probability Surveys}, 2, 191-212.

\bibitem[Boyd and Vandenberghe (2004)]{boyd}
Boyd, S. P., and Vandenberghe, L. (2004). \textit{Convex optimization}. Cambridge university press.

\bibitem[Chen and Basu and Shi (2023)]{Chen_et_al_BNP}
Chen, R., Basu, S. and Shi, Q. (2023). Restricted mean survival time estimation using Bayesian nonparametric dependent mixture models. \textit{arXiv stat.ME},

\bibitem[Cont and Tankov (2004)]{tankbook}
Cont, R., and Tankov, P., (2004), \textit{Financial
modelling with jump processes}, Boca Raton, FL: Chapman \& Hall/CRC.


\bibitem[Doksum (1974)]{doksum}
Doksum, K. (1974). Tailfree and neutral random probabilities and their posterior distributions. \textit{Ann. Probab.} \textbf{2}, 183--201.

\bibitem[Dykstra and Laud (1981)]{dykstralaud}
Dykstra, R.L. and Laud, P. (1981). A Bayesian nonparametric approach to reliability. \textit{Ann. Statist.} \textbf{9}, 356--367.

\bibitem[Epifani, Lijoi and Pr\"unster (2003)]{epilijoiprunst}
Epifani, I., Lijoi, A., and Pr\"unster, I. (2003). Exponential functionals and means of neutral‐to‐the‐right priors. \textit{Biometrika}, 90(4), 791-808.

\bibitem[Epifani and Lijoi (2010)]{epilijoi} Epifani, I. and Lijoi. A. (2010). Nonparametric priors for vectors of survival functions. \textit{Statistica Sinica}, \textbf{20.4}, 1455--1484.

\bibitem[Ferguson (1974)]{ferguson1974} Ferguson, T.S. (1974). Prior distributions on spaces of probability measures. \textit{Ann. Statist.} \textbf{2}, 615--629.

\bibitem[Ghosal and Van der Vaart (2017)]{ghosalvandervaart}
Ghosal, S., and Van der Vaart, A. (2017). \textit{Fundamentals of nonparametric Bayesian inference}. Cambridge University Press.

\bibitem[Goldman et al. (1996)]{aids}
Goldman, A., Carlin, B., Crane, L., Launer, C., Korvick, J., Deyton, L. and Abrams, D. (1996) Response of CD4+ and clinical consequences to treatment using ddI or ddC in patients with advanced HIV infection. \textit{Journal of Acquired Immune Deficiency Syndromes and Human Retrovirology}, 11,
161–169.

\bibitem[Griffin and Leisen (2017)]{GL2017} 
Griffin J. E.  and Leisen F. (2017), 'Compound random measures and their use in Bayesian nonparametrics', \textit{Journal of the Royal Statistical Society - Series B}, \textbf{79}, 525-545.

\bibitem[Hirsch and Yor (2013)]{hirschyor}
Hirsch, F., and Yor, M. (2013). On the Mellin transforms of the perpetuity and the remainder variables associated to a subordinator. \textit{Bernoulli}, 19(4), 1350-1377.

\bibitem[Hjort (1990)]{hjort}
Hjort, N.L. (1990). Nonparametric Bayes estimators based on beta processes in models for life history data. \textit{Ann. Statist.} \textbf{18}, 1259-1294.

\bibitem[Hyndman (1996)]{hyndman}
Hyndman, R. J. (1996). Computing and graphing highest density regions. \textit{The American Statistician}, 50(2), 120-126.

\bibitem[Irwin (1949)]{irwin}
Irwin, J. O. (1949). The standard error of an estimate of expectation of life, with special reference to expectation of tumourless life in experiments with mice. \textit{Epidemiology \& Infection}, 47(2), 188-189.

\bibitem[Jaynes (2003)]{jaynes}
Jaynes, E. T. (2003). \textit{Probability theory: The logic of science}. Cambridge university press.

\bibitem[Karrison (1987)]{karrison}
Karrison, T. (1987). Restricted mean life with adjustment for covariates. \textit{Journal of the American Statistical Association}, 82(400), 1169-1176.

\bibitem[Kim (2017)]{kim}
Kim, G. (2017). Bayesian test for the differences of survival functions in multiple groups. \textit{Communications for Statistical Applications and Methods}, 24(2), 115-127.


\bibitem[Lawrence et al. (2019)]{lawrence}
Lawrence, J., Qiu, J., Bai, S., and Hung, H. J. (2019). Difference in Restricted Mean Survival Time: Small Sample Distribution and Asymptotic Relative Efficiency. \textit{Statistics in Biopharmaceutical Research}, 11(1), 61-66.

\bibitem[Lee et al. (2018)]{lee}
Lee, C. H., Ning, J., and Shen, Y. (2018). Analysis of restricted mean survival time for length‐biased data. \textit{Biometrics}, 74(2), 575-583.

\bibitem[Lo (1984)]{lo}
Lo, A.Y. (1984). On a class of Bayesian nonparametric estimates: I. Density estimation. \textit{Ann. Statist.} \textbf{12}, 351--357.

\bibitem[Mantel (1966)]{mantel}
Mantel, N. (1966). Evaluation of survival data and two new rank order statistics arising in its consideration. \textit{Cancer Chemother. Rep.}, 50, 163-170.


\bibitem[Murray (2001)]{murray}
Murray, S. (2001). Using Weighted Kaplan‐Meier Statistics in Nonparametric Comparisons of Paired Censored Survival Outcomes. \textit{Biometrics}, 57(2), 361-368.

\bibitem[Pepe and Fleming (1989)]{pepe}
Pepe, M. S., and Fleming, T. R. (1989). Weighted Kaplan-Meier statistics: a class of distance tests for censored survival data. \textit{Biometrics}, 497-507.

\bibitem[Pepe and Fleming (1991)]{pepe2}
Pepe, M. S., and Fleming, T. R. (1991). Weighted Kaplan‐Meier statistics: Large sample and optimality considerations. \textit{Journal of the Royal Statistical Society: Series B (Methodological)}, 53(2), 341-352.

\bibitem[Peto and Peto (1972)]{peto}
Peto, R., and Peto, J. (1972). Asymptotically efficient rank invariant test procedures. J\textit{ournal of the Royal Statistical Society: Series A (General)}, 135(2), 185-198.

\bibitem[Riva-Palacio and Leisen (2018)]{riva2018a}
Riva-Palacio, A., and Leisen, F. (2018). Bayesian nonparametric estimation of survival functions with multiple-samples information. \textit{Electronic Journal of Statistics}, 12(1), 1330-1357.

\bibitem[Riva-Palacio and Leisen (2021)]{riva2021}
Riva-Palacio, A., and Leisen, F. (2021). Compound vectors of subordinators and their associated positive L\'evy copulas. Journal of Multivariate Analysis, 183, 104728.

\bibitem[Riva-Palacio et al. (2022)]{riva2022}
Riva-Palacio, A., Leisen, F., and Griffin, J. (2022). Survival regression models with dependent Bayesian nonparametric priors. \textit{Journal of the American Statistical Association}, 117(539).


\bibitem[Rizopoulos (2010)]{jm}
Rizopoulos, D. (2010). JM: An R package for the joint modelling of longitudinal and time-to-event data. \textit{Journal of Statistical Software}, 35, 1-33.

\bibitem[Robbins and Monro (1951)]{RobbinsMonro}
Robbins, H., and Monro, S. (1951). A stochastic approximation method. \textit{The Annals of Mathematical Statistics}, 400-407.

\bibitem[Royston and Parmar (2011)]{roystonparmar}
Royston, P., and Parmar, M. K. (2011). The use of restricted mean survival time to estimate the treatment effect in randomized clinical trials when the proportional hazards assumption is in doubt. \textit{Statistics in Medicine}, 30(19), 2409-2421.


\bibitem[Salminen and Vostrikova (2019)]{salminen}
Salminen, P.,and Vostrikova, L. (2019). On moments of integral exponential functionals of additive processes. \textit{Statistics \& Probability Letters}, 146, 139-146.


\bibitem[Sato (2013)]{sato}
Sato, K.-I. (2013). \textit{L\'evy processes and infinitely divisible distributions}. Cambridge university press.

\bibitem[Tian et al. (2014)]{tian}
Tian, L., Fu, H., Ruberg, S. J., Uno, H., and Wei, L. J. (2018). Efficiency of two sample tests via the restricted mean survival time for analyzing event time observations. \textit{Biometrics}, 74(2), 694-702.

\bibitem[Walker and Muliere (1997)]{betastacy}
Walker, S., and Muliere, P. (1997). Beta-Stacy processes and a generalization of the P\'olya-urn scheme. \textit{The Annals of Statistics}, 1762-1780.

\bibitem[Zhao and Yin (2022)]{zhang}
Zhang, C., and Yin, G. Bayesian nonparametric analysis of restricted mean survival time. \textit{Biometrics}, 79, 1383-1396.

\bibitem[Zhao et al. (2012)]{zhao}
Zhao, L., Tian, L., Uno, H., Solomon, S. D., Pfeffer, M. A., Schindler, J. S., and Wei, L. J. (2012). Utilizing the integrated difference of two survival functions to quantify the treatment contrast for designing, monitoring, and analyzing a comparative clinical study. \textit{Clinical trials}, 9(5), 570-577.

\bibitem[Zhao et al. (2016)]{zhao2016}
Zhao, L. , Claggett, B, Tian, L., Uno, H.,  Pfeffer, M. A., Solomon, S. D., Trippa, L. and Wei, L. J. (2016). On the restricted mean survival time curve in survival analysis. \textit{Biometrics}, 72(1), 215-221.

\bibitem[Zucker (1998)]{zucker}
Zucker, D. M. (1998). Restricted mean life with covariates: modification and extension of a useful survival analysis method. \textit{Journal of the American Statistical Association}, 93(442), 702-709.

\end{thebibliography}
